\documentclass[aps,pre,twocolumn,notitlepage,nofootinbib,longbibliography]{revtex4-2}

\usepackage[T1]{fontenc}
\usepackage[utf8]{inputenc}
\usepackage{lmodern}
\usepackage{amsmath,amssymb,mathtools}
\usepackage{amsthm}
\usepackage{physics}
\usepackage{microtype}
\usepackage[colorlinks=true,citecolor=blue,linkcolor=blue,urlcolor=blue]{hyperref}

\usepackage{mathrsfs}
\usepackage{bm} 

\newcommand{\bbR}{\mathbb{R}}
\newcommand{\cM}{\mathcal{M}}
\newcommand{\cG}{\mathcal{G}}
\newcommand{\cL}{\mathcal{L}}
\newcommand{\br}{\mathbf{r}}
\newcommand{\bk}{\mathbf{k}}
\newcommand{\bp}{\mathbf{p}}

\newcommand{\Id}{\mathbf{1}}
\newcommand{\diver}{\nabla\!\cdot}
\newcommand{\diverG}{\nabla_{\Gamma}\!\cdot}
\newcommand{\avg}[1]{\left\langle #1 \right\rangle}
\newcommand{\kB}{k_{\mathrm B}}
\newcommand{\exd}{\mathfrak d}
\newcommand{\ii}{\mathrm{i}}
\begin{document}

\title{Unified Gauge--Geometry Symmetry for Equilibrium Statistical Mechanics}

\author{Hai Pham-Van}
\email{haipv@hnue.edu.vn}
\affiliation{Department of Physics, Hanoi National University of Education, 136 Xuan Thuy, Cau Giay, Hanoi, Vietnam}
\affiliation{Institute of Natural Sciences, Hanoi National University of Education, 136 Xuan Thuy, Cau Giay, Hanoi, Vietnam}

\begin{abstract}

We present a symmetry-based framework for equilibrium statistical mechanics that formulates a single Lie group combining conventional spacetime symmetries with a recently identified phase-space gauge-shifting invariance [M\"uller~\textit{et~al.}, Phys.~Rev.~Lett.~\textbf{133}, 217101 (2024)]. Using Noether’s theorem, we obtain a set of general Ward identities together with previously unexplored cross-relations arising from the noncommutation of different symmetry generators. The approach extends standard many-body symmetries--such as translations, rotations, Galilean boosts, dilations, and particle exchange--by incorporating an internal gauge-shift symmetry within a unified group structure. The resulting Lie algebra suggests a hierarchy of exact identities that encompass established sum rules and indicate possible cross-coupling relations between distinct response and correlation functions. We also identify a Wigner–Eckart–Ward reduction that simplifies tensor–hyperforce correlators to two scalar radial spectra in isotropic fluids, and we outline an equivariant gauge-constrained-DFT formulation whose Euler–Lagrange equations are constructed to satisfy the corresponding Ward and cross-Ward constraints. This framework provides a consistent organizational basis for phenomena in liquids, mixtures, and interfaces, and may offer a symmetry-based perspective connecting structure, mechanics, and dynamics in many-body systems.
\end{abstract}

\maketitle

\section{Introduction}
\label{sec:intro}

Gauge invariance plays a foundational role across physics, linking symmetries to conservation laws via Noether’s theorem~\cite{banados2016short,kosmann2010noether}. In quantum field theory, local gauge invariance imposes fundamental constraints represented by Yang–Mills charge conservation~\cite{yang1954conservation,peskin2018introduction}. Gauge concepts are usually developed in the context of fields and fundamental forces~\cite{yang1954conservation,weinberg1995quantum}, yet recent work shows analogous structures in many-body statistical mechanics via a phase-space `shifting' symmetry that leaves equilibrium averages unchanged. Schmidt and co-workers applied Noether’s calculus to classical statistical functionals, deriving exact balance identities (sum rules) for interparticle forces and introducing higher-order “hyperforce” fields that generalize force correlations to arbitrary observables~\cite{hermann2024noether,robitschko2024hyperforce,muller2024gauge}. These results first emerged in general Noether form and were then re-formulated in a fully gauge-invariant way, with the shifting symmetry’s generators closing a noncommutative Lie algebra. Notably, even finite yet non-infinitesimal shifts in phase space have been shown to leave all equilibrium observables invariant, establishing a genuine statistical-mechanical gauge symmetry that generates systematic sum rules and correlation constraints, including those that apply to multi-species mixtures~\cite{Matthes2025Gauge}. This form of thermal gauge invariance provides a new route for uncovering conservation-law analogues in equilibrium many-body systems. It yields not only the standard force-balance relations but also hyperforce sum rules that apply to arbitrary phase-space observables, in a manner reminiscent of Hirschfelder’s classical hypervirial theorem~\cite{hermann2021noether, robitschko2024hyperforce,hirschfelder1960classical}.   

Classical and quantum many-body systems host a rich spectrum of symmetries, including spatial translations, rotations, and scale transformations, as well as time translations, Galilean boosts, particle-permutation or species-exchange operations, and symmetries associated with order parameters. Each of these symmetries, according to Noether’s principle, gives rise to a corresponding conservation law or exact identity within the system~\cite{banados2016short}. Combining distinct symmetries has historically yielded many central results in physics. For example, in equilibrium liquids the Euclidean spatial invariances (translations+rotations, and uniform scaling) link structural correlations to thermodynamic responses via the compressibility and virial routes~\cite{bradlyn2012kubo,hansen2013theory,irving1950statistical}. The translational symmetry enforces the compressibility sum rule relating the static structure factor to the isothermal compressibility, while rotational and Galilean invariance underlie the Irving–Kirkwood virial theorem connecting microscopic forces to pressure~\cite{hansen2013theory,irving1950statistical,kirkwood1951statistical}.

 Space–time translations combined with Galilean boosts enforce Ward or Slavnov–Taylor identities for hydrodynamics and Kubo relations for viscosity and conductivity~\cite{bradlyn2012kubo,kadanoff1963hydrodynamic}. Adding scale to translations and rotations in conformal or Schrödinger settings fixes correlation functions and dispersion relations with little freedom~\cite{polyakov1970conformal}. Rigid motions together with additivity support morphometric thermodynamics, which reduces interfacial free energies to four geometric invariants. Translational invariance with charge neutrality gives the Stillinger–Lovett moment conditions for electrolytes~\cite{martin1988sum}. Time translation with microreversibility yields fluctuation theorems such as Jarzynski and Crooks~\cite{jarzynski1997nonequilibrium,crooks1999entropy}. What remains absent is a single unified symmetry group that simultaneously incorporates all conventional symmetries together with the additional phase-space gauge shifting. Earlier formulations addressed each symmetry, or at most pairs of symmetries, in isolation, and no general framework has existed that integrates geometric, dynamical, internal, and gauge-like invariances within a coherent theoretical structure.

In this work, we embed spatial symmetries--translational, rotational, and scaling--together with dynamical, compositional, and phase-space gauge-shifting symmetries into a single Lie-group structure. By analyzing the noncommuting elements of its Lie algebra, we obtain a series of cross-symmetry identities that extend existing relations between different symmetry sectors. The resulting equivariant framework suggests several connections that may link distinct aspects of equilibrium physics. Among these, a Wigner–Eckart–Ward reduction appears to yield selection rules that simplify tensorial and higher-rank hyperforce correlation functions into scalar forms under combined symmetry operations, in close analogy with the reduction of tensor operators in the Wigner–Eckart theorem~\cite{rose1995elementary,edmonds1996angular}. The same framework also provides exact scattering-to-stress relations, leading to sum-rule identities that connect static two-point structure factors with stress correlations and viscoelastic responses. These relations may allow estimation of elastic and viscous properties without relying on direct stress autocorrelation measurements.

In addition, we propose an equivariant gauge-constrained-DFT variational principle, conceived as an extension of classical density functional theory, whose Euler–Lagrange equations are constructed to satisfy the derived Ward identities and cross-Ward constraints. Within this formulation, equilibrium densities and order-parameter profiles are expected to respect the full set of symmetry-induced sum rules, promoting internal consistency between structure and thermodynamics. Established results such as the compressibility relation, virial pressure expression, and known hyperforce balance laws are recovered as limiting cases of the general framework.

\section{Theoretical calculation}
\label{sec:mod}

Lie theory and differential geometry are familiar from quantum field
theory, yet they are less often discussed explicitly in statistical mechanics. A Lie algebra is the infinitesimal counterpart of a continuous symmetry group. Its elements represent symmetry generators, and their Lie bracket records how two infinitesimal transformations compose, thereby capturing the noncommutativity structure of the symmetry. Semidirect products provide a convenient organization when distinct symmetry sectors do not decouple, because one sector
acts on another and induces characteristic cross-relations between generators. We also use differential forms and the exterior derivative to express gradient and curl constraints and to simplify boundary
manipulations via Stokes and Gauss theorems. Exact constraints implied by symmetry are Ward identities, and a field-theoretic prototype is the
Ward--Takahashi identity. In gauge and global-symmetry settings, the
Ward--Takahashi identity expresses the invariance of the generating functional under an infinitesimal symmetry and thus relates correlation functions of currents and fields. In the present equilibrium context, the same logic yields exact relations among ensemble averages and correlators that follow solely from
symmetry and measure invariance. For further reading see
Refs.~\cite{hall2013lie,lee2003smooth,nakahara2018geometry,cardy1996scaling,zinn2021quantum}.

Consider a classical many-body system in $n$ spatial dimensions.
For a fixed particle number $N$, the phase space is
$\cM_N=(\bbR^n\times\bbR^n)^N$, with microstate (phase-space point)
$\Gamma=(\br^N;\bp^N)\in\cM_N$ and Liouville measure $d\Gamma=d\br^N\,d\bp^N$.
Here $\br^N\equiv(\br_1,\ldots,\br_N)$ and $\bp^N\equiv(\bp_1,\ldots,\bp_N)$ collect the particle
positions $\br_i\in\bbR^n$ and canonical momenta $\bp_i\in\bbR^n$.

The $N$-particle Hamiltonian is
\begin{equation}
H_N(\Gamma)
=\sum_{i=1}^N\frac{\bp_i^2}{2m_i}+u(\br^N)+\sum_{i=1}^N V_{\rm ext}(\br_i),
\end{equation}
where $m_i$ is the mass of particle $i$, $u(\br^N)$ is the many-body interaction potential, and
$V_{\rm ext}(\br_i)$ is a one-body external potential acting on particle $i$.

In the grand-canonical ensemble the full phase space is the disjoint union
$\cM=\bigsqcup_{N\ge0}\cM_N$. The grand partition function is
\begin{equation}
\Xi
=\sum_{N\ge0}\int_{\cM_N}\! \dd\Gamma\;
\exp\!\big[-\beta\big(H_N(\Gamma)-\mu N\big)\big],
\label{eq:grandXi}
\end{equation}
and the corresponding grand-canonical Gibbs weight on the $N$-particle sector is
\begin{equation}
\mathcal P_N(\Gamma)
=\Xi^{-1}\exp\!\Big[-\beta\big(H_N(\Gamma)-\mu N\big)\Big],
\label{eq:PN}
\end{equation}
where $\mu$ is the chemical potential and $\beta\equiv 1/(\kB T)$ with $\kB$ the Boltzmann constant and
$T$ the absolute temperature.

Thermal averages of a phase-space observable $A_N(\Gamma)$ are then
\begin{equation}
\avg{A}
=\sum_{N\ge0}\int_{\cM_N}\! \dd\Gamma\;\mathcal P_N(\Gamma)\,A_N(\Gamma).
\label{eq:grandavg}
\end{equation}

If the particle number is fixed (canonical ensemble), the same formulas apply with the sum over $N$
removed. The normalization is the canonical partition function
\begin{equation}
Z_N \equiv \int_{\cM_N} \dd\Gamma\,e^{-\beta H_N(\Gamma)},
\end{equation}
and thermal averages are
\begin{equation}
\avg{A}_N
=Z_N^{-1}\int_{\cM_N} \dd\Gamma\,e^{-\beta H_N(\Gamma)}A_N(\Gamma).
\end{equation}

The guiding idea of what follows is simple: once we know how a symmetry acts on microscopic variables,
Liouville-volume preservation together with integration by parts under the Gibbs weight produces identities
that relate the symmetry variation of observables to that of the energy.

\subsection{Unified symmetry group}
\label{subsec:symgroup}

We introduce a unified symmetry group $\cG$ that collects geometric–time kinematics, internal symmetries, and discrete involutions:
\begin{equation}
\cG \;=\; \Big(\,\bbR_D \ltimes \mathrm{Gal}(n)\,\Big)\;\ltimes\;
\Big(\,\cG_{\rm sh} \times \cG_{\rm sp} \times \cG_{\rm ord}\,\Big)\;\ltimes\;
\Big(\,\mathbb{Z}_2^{\mathsf T}\times \mathbb{Z}_2^{\mathsf P}\,\Big).
\label{eq:unified-group}
\end{equation}
Here $\mathrm{Gal}(n)$ is the Galilei group generated by spatial translations, rotations, Galilean boosts, and time translations; the dilation factor $\bbR_D$ implements uniform spatial scaling and acts by automorphisms on $\mathrm{Gal}(n)$. The internal sector combines the phase–space gauge–shift group $\cG_{\rm sh}$, the species symmetry $\cG_{\rm sp}$ (permutations of components or continuous mixing, depending on the model), and the order–parameter symmetry $\cG_{\rm ord}$ (for example $O(n)$ for vector order, $U(1)$ for phase fields, or a tensor representation). The final factor $\mathbb{Z}_2^{\mathsf T}$ and $\mathbb{Z}_2^{\mathsf P}$ collects the discrete symmetries of time reversal $\mathsf T$ and parity $\mathsf P$; they act by conjugation on all other factors and constrain correlation functions even though they are not connected Lie symmetries. 
The symbol $\times$ denotes a direct product, while the symbol $\ltimes$ denotes a semidirect product.

The infinitesimal counterpart of the group $\cG$ is its Lie algebra
$\mathfrak g\equiv \mathrm{Lie}(\cG)=T_e\cG$, i.e.\ the tangent space at the identity element $e$ of the connected component.
Each generator $X\in\mathfrak g$ defines a one-parameter subgroup $g_s=\exp(sX)\in\cG$.
We work primarily at this Lie-algebra level because the Stein identities below follow from integration by parts applied to the corresponding infinitesimal phase-space flows.
The discrete factors $\mathbb{Z}_2^{\mathsf T}$ and $\mathbb{Z}_2^{\mathsf P}$ are not connected to $e$ and therefore do not contribute generators to $\mathfrak g$, although they constrain correlators by conjugation.
The semidirect products in Eq.~\eqref{eq:unified-group} represent how different symmetry sectors act on each other; at the Lie-algebra level this yields nonvanishing commutators and is the origin of the cross-Ward identities derived below.
The gauge-shift factor $\cG_{\rm sh}$ may be regarded as a suitable (in general infinite-dimensional) subgroup of the diffeomorphism group of space;
its Lie algebra consists of smooth vector fields $\bm\epsilon(\br)$ and forms a subalgebra $\mathfrak g_{\rm sh}\subset\mathfrak g$.
For background on Lie groups, Lie derivatives, and related geometric notions, see e.g.\ Refs.~\cite{hall2013lie,lee2003smooth,nakahara2018geometry}.

We use a (left) representation of $\mathfrak g$ by phase-space vector fields on $\cM$:
to each generator $X\in\mathfrak g$ we associate a vector field $v_X$ on $\cM$, whose flow is the infinitesimal transformation of the microscopic variables generated by $X$.
The induced action on an observable $A(\Gamma)$ is the Lie derivative
\begin{equation}
\cL_X A \equiv v_X\!\cdot\!\nabla_\Gamma A,
\end{equation}

and the representation map is a Lie-algebra homomorphism:
$[v_X,v_Y]=v_{[X,Y]}$, equivalently $[\cL_X,\cL_Y]=\cL_{[X,Y]}$.

Here $\nabla_\Gamma=(\nabla_{\br^N},\nabla_{\bp^N})$ is the full phase-space gradient with respect to
$\Gamma=(\br^N;\bp^N)$, with $\nabla_{\br^N}=(\nabla_{\br_1},\ldots,\nabla_{\br_N})$ and
$\nabla_{\bp^N}=(\nabla_{\bp_1},\ldots,\nabla_{\bp_N})$.
We reserve $\nabla$ for the spatial gradient with respect to its displayed argument,
and write $\nabla_{\br_i}$ and $\nabla_{\bp_i}$ for derivatives with respect to particle coordinates and momenta.

A central ingredient is the gauge-shift generator $G[\bm\epsilon]\in\mathfrak g$,
parametrized by a smooth spatial vector field $\bm\epsilon(\br)$.
Under the representation map $X\mapsto v_X$ it is represented on phase space by the cotangent-lift vector field
\begin{equation}
v_{G[\bm\epsilon]} \equiv v_G
=\sum_{i=1}^N\Big[\bm\epsilon(\br_i)\!\cdot\!\nabla_{\br_i}
-\big(\nabla\bm\epsilon(\br_i)\,\bp_i\big)\!\cdot\!\nabla_{\bp_i}\Big],
\label{eq:vG}
\end{equation}
where $(\nabla\bm\epsilon)_{ij}=\partial_{r_j}\epsilon_i$ is the Jacobian of $\bm\epsilon$
(so $(\nabla\bm\epsilon(\br_i)\,\bp_i)_\alpha=(\partial_{r_\beta}\bm\epsilon_\alpha(\br_i))\,p_{i\beta}$).

Equation~\eqref{eq:vG} is the standard cotangent-lift canonical form of the local phase-space shifting symmetry;
it is equivalent to the global shifting generator in Ref.~\cite{muller2024gauge} [Eq.~(7) therein], rewritten here as a phase-space vector field.
Thus $v_G$ is simply the phase-space representative of the Lie-algebra element $G[\bm\epsilon]$.
When no confusion arises, we identify a generator with its representative and write $G[\bm\epsilon]$ for $v_{G[\bm\epsilon]}$.

A straightforward calculation shows that $v_{G}$ preserves the Liouville phase-space volume,
i.e.\ $\diverG v_{G}=0$. However, when weighted by the Gibbs density $\mathcal P$, one finds a
nonzero weighted divergence

\begin{equation}
\diverG\!\big(\mathcal P\, v_{G}\big)
=-\beta\,\mathcal P\,v_{G}\!\cdot\!\nabla_\Gamma H,
\label{eq:weighted-div}
\end{equation}
with $H\equiv H_N(\Gamma)$ (or $H-\mu N$ in the grand ensemble) understood as the energy entering the Gibbs weight.

Assuming sufficient regularity of $A$ and vanishing boundary terms, integration by parts under the Gibbs
measure implies the finite-temperature Stein identity~\cite{LeySwan2013} for any generator
$X\in\mathfrak g$ whose representative flow is volume-preserving ($\diverG v_X=0$):
\begin{equation}
\avg{\cL_X A}=\beta\,\avg{A\,\cL_X H}.
\label{eq:thermal-ward}
\end{equation}
Here $\avg{\cdot}$ denotes the thermal ensemble average (canonical or grand-canonical) with respect to
$\mathcal P$. Equivalently, Eq.~\eqref{eq:thermal-ward} can be written as
$\avg{\cL_X\big(A\,e^{-\beta H}\big)}=0$.

Equation~\eqref{eq:thermal-ward} follows from integration by parts under the Gibbs weight and therefore
assumes that the phase-space domain has no boundary (as for periodic boundary conditions) or that the
associated surface term vanishes. More generally, let the accessible configuration space for $N$ particles
be a region $\Omega_N\subset(\bbR^{n})^{N}$ with boundary $\partial\Omega_N$
(e.g.\ $\Omega_N=(\{z>0\})^{N}$ for a planar hard wall). For a volume-preserving generator, the same calculation yields the boundary Stein identity
\begin{equation}
\begin{aligned}
\avg{\cL_X A}
={}&\beta\,\avg{A\,\cL_X H}
\\
&+\frac{1}{Z_N}
\oint_{\partial\Omega_N}\! \dd S_{\br^N}
\int \dd \bp^N\;
A(\br^N,\bp^N)\,e^{-\beta H_N(\br^N,\bp^N)}
\\
&\hspace{3em}\times
v_X^{\,\br}(\br^N,\bp^N)\!\cdot\!\mathbf n .
\end{aligned}
\label{eq:thermal-ward-boundary}
\end{equation}
where $v_X^{\,\br}$ denotes the configuration-space part of $v_X$ and $\mathbf n$ is the outward unit normal
on $\partial\Omega_N$. Equation~\eqref{eq:thermal-ward} is recovered whenever this surface contribution
vanishes, which holds for periodic boundaries, for sufficiently confining soft potentials, or for
generators whose flows are tangent to the boundary ($v_X^{\,\br}\!\cdot\!\mathbf n=0$ on $\partial\Omega_N$).
When a boundary breaks a symmetry, keeping the surface term in
Eq.~\eqref{eq:thermal-ward-boundary} produces boundary Ward identities; Eq.~\eqref{eq:C4} is an example for a planar wall.

\subsection{Lie algebra: nonzero commutators}
\label{subsec:lie}

We denote $T(\mathbf a)$ as the generator of spatial translations  
by the constant vector $\mathbf a$; $R(\Theta)$ is the rotation generator with antisymmetric  
matrix $\Theta$ acting on $\mathbb{R}^n$; $D$ is the (spatial) dilation generator;  
$\mathcal{B}(\mathbf u)$ is the Galilean boost generator with constant velocity $\mathbf u$;  
$t(\tau)$ is the time‑translation generator by the amount $\tau$;    
and $\Pi[\bm\chi]$ is the momentum–fiber translation generated by the smooth field $\bm\chi(\br)$.  
The symbols $\bm\epsilon_{1,2}$ denote two independent gauge fields.  
The composite field $\bm\chi_{\bm\epsilon,\mathbf u}$ that appears in the boost–gauge bracket  
is defined particlewise, for a particle of mass $m_i$, by
\begin{equation}
\bm\chi_{\bm\epsilon,\mathbf u}(\br_i)
= m_i\,(\mathbf u\cdot\nabla)\bm\epsilon(\br_i)
= (\nabla\bm\epsilon(\br_i))\,(m_i\mathbf u).
\end{equation}

The explicit forms of the operators are given by
\begin{align}
T(\mathbf a)
&=\sum_{i=1}^N \mathbf a\!\cdot\!\nabla_{\br_i},
\label{eq:vT}\\
R(\Theta)
&=\sum_{i=1}^N\Big[(\Theta\br_i)\!\cdot\!\nabla_{\br_i}
+(\Theta\bp_i)\!\cdot\!\nabla_{\bp_i}\Big],
\label{eq:vR}\\
D
&=\sum_{i=1}^N\Big[\br_i\!\cdot\!\nabla_{\br_i}
-\bp_i\!\cdot\!\nabla_{\bp_i}\Big],
\label{eq:vD}\\
\mathcal{B}(\mathbf u)
&=\sum_{i=1}^N\Big[t\,\mathbf u\!\cdot\!\nabla_{\br_i}
+m_i\mathbf u\!\cdot\!\nabla_{\bp_i}\Big],
\label{eq:vB}\\
t(\tau)
&=\tau\,\partial_t,
\label{eq:vtgen}\\
\Pi[\bm\chi]
&=\sum_{i=1}^N \bm\chi(\br_i)\!\cdot\!\nabla_{\bp_i}.
\label{eq:vPi}
\end{align}

The Lie bracket of phase-space vector fields $U,V$ representing the generators is
\begin{equation}
[U,V]= (U\!\cdot\!\nabla_\Gamma)\,V-(V\!\cdot\!\nabla_\Gamma)\,U,
\end{equation}
while for spatial vector fields $\bm\epsilon_1,\bm\epsilon_2$ we use
\begin{equation}
[\bm\epsilon_1,\bm\epsilon_2]
=(\bm\epsilon_1\!\cdot\!\nabla)\bm\epsilon_2-(\bm\epsilon_2\!\cdot\!\nabla)\bm\epsilon_1.
\end{equation}

The nonzero commutators involving the gauge generators are (see App.~\ref{app:commutators} for a detailed derivation):
\begin{align}
[T(\mathbf a),\,G[\bm\epsilon]] 
&= G\!\left[(\mathbf a\cdot\nabla)\bm\epsilon\right],
\label{eq:comm-TG} \\[3pt]
[R(\Theta),G[\bm\epsilon]]
&=G\!\left[(\Theta\br\!\cdot\!\nabla)\bm\epsilon-(\Theta\,\bm\epsilon)\right].
\label{eq:comm-RG} \\[3pt]
[D,\,G[\bm\epsilon]] 
&= G\!\left[(\br\cdot\nabla)\bm\epsilon - \bm\epsilon\right],
\label{eq:comm-DG} \\[3pt]
[\mathcal{B}(\mathbf u),\,G[\bm\epsilon]] 
&= G\!\left[t\,(\mathbf u\cdot\nabla)\bm\epsilon\right]
+ \Pi\!\left[-\,\bm\chi_{\bm\epsilon,\mathbf u}\right],
\label{eq:comm-BG} \\[3pt]
[G[\bm\epsilon_1],\,G[\bm\epsilon_2]] 
&= G\!\left[[\bm\epsilon_1,\bm\epsilon_2]\right],
\label{eq:comm-GG} \\[3pt]
[\Pi[\bm\chi],\,G[\bm\epsilon]]
&= \Pi\!\left[-\,(\bm\epsilon\cdot\nabla)\bm\chi 
- (\nabla\bm\epsilon)\,\bm\chi\right].
\label{eq:comm-PiG}
\end{align}
\vspace{-5ex}
\begin{align}
[t(\tau),\,G[\bm\epsilon]] 
&= 
\begin{cases}
0, & \text{if } \bm\epsilon \text{ is time-independent}, \\
G[\tau\,\partial_t\bm\epsilon], & \text{otherwise}.
\end{cases}
\label{eq:comm-tG}
\end{align}

In what follows we use the Stein identity~\eqref{eq:thermal-ward} to derive the constraints collected in
Sec.~\ref{sssec:gauge-u1-master}. For any generators $X,Y\in\mathfrak g$, the phase-space representation
$X\mapsto v_X$ is a Lie-algebra homomorphism, so the induced Lie derivatives satisfy
\begin{equation}
[\cL_X,\cL_Y]=\cL_{[X,Y]},
\end{equation}
with $[X,Y]\in\mathfrak g$ the Lie bracket. Consequently, each nonvanishing commutator in
Eqs.~\eqref{eq:comm-TG}--\eqref{eq:comm-tG} yields a cross-correlation constraint by applying
Eq.~\eqref{eq:thermal-ward} to the bracket generator $[X,Y]$. For generators parametrized by arbitrary
test fields, this Ward identity localizes and produces the correlator relations given below, in direct
analogy with Ward--Takahashi/Schwinger--Dyson mechanisms in field theory~\cite{peskin2018introduction,zinn2021quantum}.

\subsection{Gauge master identity and cross-Ward constraints}
\label{sssec:gauge-u1-master}

\subsubsection{Local gauge shift \texorpdfstring{$\times$}{x} global \texorpdfstring{$U(1)$}{U(1)}}
\label{sssec:local-gaugeshift-x-u1}

Introduce the microscopic density
\(\hat\rho(\mathbf r)=\sum_i\delta(\mathbf r-\mathbf r_i)\)
and internal-force density
\(\hat{\mathbf F}_{\mathrm{int}}(\mathbf r)=\sum_{i<j}\mathbf f_{ij}\,[\delta(\mathbf r-\mathbf r_i)-\delta(\mathbf r-\mathbf r_j)]\)
with pair force \(\mathbf f_{ij}\), where \(\delta(\cdot)\) is the Dirac delta function.
For any configuration-only observable \(B(\mathbf r^N)\) define the cross-correlators
\begin{equation}
\mathbf C_{B F_{\mathrm{int}}}(\mathbf r)
\equiv \big\langle B\,\hat{\mathbf F}_{\mathrm{int}}(\mathbf r)\big\rangle,
\qquad
C_{B\rho}(\mathbf r)\equiv \big\langle B\,\hat\rho(\mathbf r)\big\rangle .
\end{equation}

In a translationally invariant bulk fluid ($V_{\rm ext}=0$), the one-point averages
$\langle \hat\rho(\br)\rangle=\rho$ and $\langle \hat{\mathbf F}_{\rm int}(\br)\rangle=\mathbf 0$
are constants. The mixed correlators $C_{B\rho}(\br)$ and $\mathbf C_{B F_{\rm int}}(\br)$ are
nevertheless nontrivial for local or translation-covariant insertions $B$ (e.g.\
$B=\delta\hat\rho(\mathbf 0)$ or $B=\delta\hat\rho_{-\bk}$): translational invariance then implies
dependence only on the separation from the insertion point, i.e.\ on $\br$ (or $\bk$).
Only for translationally invariant $B$ (in particular $B=1$) do these correlators collapse to
one-point averages and the bulk identity reduces to $0=0$.

The local spatial gauge-shift symmetry \(G[\bm\epsilon]=\sum_i \bm\epsilon(\mathbf r_i)\!\cdot\!\nabla_{\mathbf{r}_i}\),
together with global \(\mathrm{U}(1)\) particle-number invariance, implies an exact equilibrium identity.
In bulk (\(V_{\rm ext}=0\)) and for configuration observables \(B(\br^N)\), and upon the standard
separation of contact terms at the insertion point, this identity reduces to the gauge master identity
\begin{align}
\mathbf C_{B F_{\mathrm{int}}}(\mathbf r)
&= k_B T\,\nabla C_{B\rho}(\mathbf r),
\label{eq:gauge-master-real}\\
\mathrm{i}\,\mathbf k\!\cdot\!
\mathbf C_{B F_{\mathrm{int}}}(\mathbf k)
&= -k_B T\,k^2\,C_{B\rho}(\mathbf k),
\label{eq:gauge-master-fourier}
\end{align}
where \(\mathbf k\) is the wavevector conjugate to \(\mathbf r\).

Equation~\eqref{eq:gauge-master-real} is the bulk/contact-term specialization of the
hyperforce-balance identity generated by the same local shifting symmetry~\cite{robitschko2024hyperforce,Matthes2025Gauge}.
In its most general form this hyperforce sum rule holds for arbitrary phase-space observables
\(A(\br^N,\bp^N)\) and reads
\begin{equation}
\begin{aligned}
\avg{A\,\hat{\mathbf F}(\br)}
&=
-\kB T\,\Big\langle
\sum_i \delta(\br-\br_i)\,\nabla_{\mathbf{r}_i} A
\Big\rangle
\\
&\quad
-\kB T\,\nabla\!\cdot\!\Big\langle
\sum_i \delta(\br-\br_i)\,
\nabla_{\bp_i}A\otimes\bp_i
\Big\rangle .
\end{aligned}
\label{eq:hyperforce-phase-space}
\end{equation}
where $\otimes$ is the dyadic (outer) product,
$(\mathbf a\otimes\mathbf b)_{\alpha\beta}=a_\alpha b_\beta$, the total force density
\(
\hat{\mathbf F}(\br)=\nabla\!\cdot\!\hat{\boldsymbol\tau}(\br)
+\hat{\mathbf F}_{\mathrm{int}}(\br)
-\hat\rho(\br)\nabla V_{\rm ext}(\br)
\)
and kinematic stress
\(
\hat{\boldsymbol\tau}(\br)\equiv -\sum_i\delta(\br-\br_i)\,\bp_i\otimes\bp_i/m_i
\).
For configuration observables \(B(\br^N)\) the momentum-flux term vanishes, and separating internal and
external contributions gives
\begin{equation}
\begin{aligned}
\avg{B\,\hat{\mathbf F}_{\mathrm{int}}(\br)}
&-\avg{B\,\hat\rho(\br)}\,\nabla V_{\rm ext}(\br)
\\
&=
\kB T\,\nabla \avg{B\,\hat\rho(\br)}
\\
&\quad
-\kB T\,\Big\langle
\sum_i \delta(\br-\br_i)\,
\nabla_{\mathbf r_i} B
\Big\rangle .
\end{aligned}
\label{eq:connection-hyperYBG}
\end{equation}

Choosing $B=1$ in Eq.~\eqref{eq:connection-hyperYBG} yields the standard
Yvon--Born--Green (first BBGKY) equilibrium force-balance relation,
\(
\langle \hat{\mathbf F}_{\rm int}(\br)\rangle-\rho(\br)\nabla V_{\rm ext}(\br)=\kB T\,\nabla\rho(\br),
\)
which becomes $0=0$ for a homogeneous bulk fluid ($V_{\rm ext}=0$, $\rho=\mathrm{const.}$).

For the translation-covariant choices used below (e.g.\ \(B=\delta\hat\rho(\mathbf 0)\)),
the last term in Eq.~\eqref{eq:connection-hyperYBG} is a local contact contribution at the insertion point
and it only fixes the self part of \(C_{B\rho}\); for \(\br\neq\mathbf 0\) one recovers the compact
bulk form in Eq.~\eqref{eq:gauge-master-real}.
Under confinement one retains the \(V_{\rm ext}\) term in Eq.~\eqref{eq:connection-hyperYBG} and works in
real space (or along the surviving translational directions).
In the remainder of this work we take Eq.~\eqref{eq:gauge-master-real} as the starting point, because
combining the gauge generator with rotations, dilations, boundaries, and boosts in the unified algebra
yields the cross-Ward constraints derived below.

\subsubsection{Gauge \texorpdfstring{$\times$}{x} \texorpdfstring{$U(1)$}{U(1)}: divergence form for the density choice}
\label{sssec:gauge-u1-divergence}

Choosing $B=\delta\hat\rho(\mathbf 0)$—with $\hat\rho(\mathbf 0)=\sum_i \delta(\mathbf r_i)$ and $\delta\hat\rho(\mathbf r)\equiv \hat\rho(\mathbf r)-\rho$ the density fluctuation—and writing the pair structure as $g(\mathbf r)=1+h(\mathbf r)$, with $h$ the total correlation function, and the static structure factor as $S(\mathbf k)=1+\rho\,\tilde h(\mathbf k)$ where $\tilde h(\mathbf k)=\int \mathrm{d}\mathbf r\,e^{-\mathrm{i}\mathbf k\cdot\mathbf r}\,h(\mathbf r)$, and where $\rho\equiv\langle \hat\rho(\mathbf r)\rangle$ is the bulk number density, the same gauge–$\mathrm U(1)$ identity yields the divergence form

\begin{equation}
\begin{aligned}
\nabla\!\cdot 
\mathbf C_{\delta\rho F_{\mathrm{int}}}(\mathbf r)
&= k_B T\,\rho^2\,\nabla^2 h(\mathbf r), \\
\mathrm{i}\,\mathbf k\!\cdot\!
\mathbf C_{\delta\rho F_{\mathrm{int}}}(\mathbf k)
&= -k_B T\,\rho\,k^2\,[S(\mathbf k)-1],
\end{aligned}
\label{eq:C1}
\end{equation}

where
\begin{equation}
\mathbf C_{\delta\rho F_{\mathrm{int}}}(\br)
= \big\langle \delta\hat\rho(\mathbf 0)\,\hat{\mathbf F}_{\mathrm{int}}(\br)\big\rangle,
\label{eq:CrhoF-real}
\end{equation}

and its Fourier transform is
\begin{equation}
\mathbf C_{\delta\rho F_{\mathrm{int}}}(\bk)
=\int \mathrm{d}^n r\,e^{-\mathrm{i}\bk\cdot\br}\,
\mathbf C_{\delta\rho F_{\mathrm{int}}}(\br)
= V^{-1}\,\big\langle \delta\hat\rho_{-\bk}\,\hat{\mathbf F}_{\mathrm{int}}(\bk)\big\rangle,
\label{eq:CrhoF-k}
\end{equation}

with $V$ the system volume (area for $n=2$, length for $n=1$), i.e., the normalization associated with the Fourier convention used. 
Equation~\eqref{eq:C1} is complementary to the second-order Noether ``3g-rule''
of Ref.~\cite{sammuller2023noether}: since $h=g-1$, the Laplacian $\nabla^2 h$ is the trace of
$\nabla\nabla g$ in Eq.~(12) of Ref.~\cite{sammuller2023noether}.

\subsubsection{Gauge \texorpdfstring{$\times$}{x} rotations \texorpdfstring{$SO(n)$}{SO(n)}: longitudinality}
\label{sssec:gauge-rot-longitudinality}

For isotropic, central–force fluids the vector correlator is purely longitudinal, so its transverse projection vanishes and one has a closed form. Physically, this longitudinality is the cross consequence of combining the gauge shift with spatial isotropy (the $SO(n)$ sector): the gauge identity fixes the longitudinal divergence, while rotational invariance forbids any transverse component. Equivalently, the noncommutativity in Eq.~\eqref{eq:comm-RG} enforces a selection rule that leaves only the $\bk$‑parallel channel, hence 
\begin{equation}
\begin{aligned}
\mathbf P_T(\mathbf k)\,
\mathbf C_{\delta\rho F_{\mathrm{int}}}(\mathbf k)=\mathbf 0.
\\
\end{aligned}
\label{eq:C2}
\end{equation}
where \(\mathbf P_T(\mathbf k)=\mathbf I-\hat{\mathbf k}\hat{\mathbf{k}}^\top\) is the transverse projector, $\mathbf I$ is the identity tensor,  \(\hat{\mathbf k}=\mathbf k/k\) and $\hat{\mathbf{k}}^\top$ is the transpose of $\hat{\mathbf{k}}$.

\subsubsection{Gauge \texorpdfstring{$\times$}{x} dilations: first radial moment and compressibility}
\label{sssec:gauge-dilation-compressibility}

The identity is the cross-consequence of combining the gauge shift with spatial dilations through the commutator Eq.~\eqref{eq:comm-DG}. Integrating the real-space gauge master identity~\eqref{eq:gauge-master-real} by parts converts the gradient relation into its first radial moment. The second equality follows by inserting the compressibility sum rule derived from translational invariance, $S(0)=\rho k_{\mathrm B}T\kappa_T$, with $S(0)$ including the self part, where $\kappa_T$ is the isothermal compressibility.

\begin{align}
\int \mathrm{d}^n r\, \br\cdot\mathbf C_{\delta \rho F_{\mathrm{int}}}(\br)
&= -\,n\,\kB T\,\rho\,[S(\mathbf 0)-1] \label{eq:C3a} \\
&= -\,n\,(\kB T)^2\,\rho^2\,\kappa_T + n\,\kB T\,\rho. \label{eq:C3b}
\end{align}

so the $\,n\,\kB T\,\rho$ term appears when $S(0)$ includes the self part; equivalently, with the connected structure factor $S_c(0)=S(0)-1$, the integral equals $-n\,\kB T\,\rho\,S_c(0)$.

\subsubsection{Gauge \texorpdfstring{$\times$}{x} boundary geometry / broken translations: wall contact relation}
\label{sssec:gauge-boundary-contact}

Near a planar wall at $z=0$ with inward normal $+\hat{\mathbf z}$, one obtains a contact relation for the hyperforce. This relation is the cross-consequence of combining the gauge shift with boundary geometry: the wall breaks normal translations while preserving tangential Euclidean symmetry, and Gauss’s theorem converts the bulk identity into a boundary term. 

We start from the bulk divergence form
Eq.~\eqref{eq:C1}, which holds pointwise in the accessible fluid region.
For a planar hard wall the accessible domain is the half-space
$\Omega=\{(\br_\parallel,z):z>0\}$; inside $\Omega$ the external potential is constant (hence its force vanishes),
so the wall enters only through the boundary of the integration domain.

Integrate Eq.~\eqref{eq:C1} over a slab
$\Omega_L\equiv \mathcal A\times(0,L)\subset\Omega$,
with finite in-plane domain $\mathcal A\subset\bbR^{n-1}$ and $L>0$.
Applying Gauss’s theorem on both sides gives the boundary form
\begin{equation*}
\oint_{\partial\Omega_L}\! \dd S\;\mathbf n\!\cdot\!\mathbf C_{\delta \rho F_{\mathrm{int}}}(\br)
=
\kB T\,\rho^2
\oint_{\partial\Omega_L}\! \dd S\;\mathbf n\!\cdot\!\nabla h(\br),
\end{equation*}
where $\mathbf n$ is the outward unit normal on $\partial\Omega_L$.
Decomposing the boundary as
$\partial\Omega_L=(\mathcal A\times\{0\})\cup(\mathcal A\times\{L\})\cup(\partial\mathcal A\times(0,L))$
yields
\begin{equation}
\begin{aligned}
&\int_{\mathcal A}\! \dd^{\,n-1}\br_\parallel\,
\Big[
C_{\delta \rho F_{\mathrm{int}},z}(\br_\parallel,L)
- C_{\delta \rho F_{\mathrm{int}},z}(\br_\parallel,0^+)
\Big]
\\
&\quad
+\oint_{\partial\mathcal A\times(0,L)}\! \dd S\;
\mathbf n_\parallel\!\cdot\!\mathbf C_{\delta \rho F_{\mathrm{int}}}
\\[0.3em]
&=\kB T\,\rho^2
\Bigg\{
\int_{\mathcal A}\! \dd ^{\,n-1}\br_\parallel\,
\Big[
\partial_z h(\br_\parallel,L)
-\partial_z h(\br_\parallel,0^+)
\Big]
\\
&\hspace{4em}
+\oint_{\partial\mathcal A\times(0,L)}\! \dd S\;
\mathbf n_\parallel\!\cdot\!\nabla h
\Bigg\}.
\end{aligned}
\end{equation}
with $\mathbf n_\parallel$ the outward normal on the lateral surface $\partial\mathcal A\times(0,L)$.
For periodic boundary conditions parallel to the wall (or, equivalently, for large $\mathcal A$ with short-ranged correlations),
the lateral surface integrals vanish. Taking $L\to\infty$ removes the $z=L$ terms because
$\mathbf C_{\delta \rho F_{\mathrm{int}}}(\br)\to\mathbf 0$ and $\nabla h(\br)\to\mathbf 0$ far from the wall.
The only remaining contribution comes from the wall plane at $z=0$, whose outward normal is $-\hat{\mathbf z}$
(hence the appearance of the difference between $z=L$ and $z=0^+$ above). Rearranging then yields the wall contact relation

\begin{equation}
\begin{aligned}
\int_{\mathcal A} \mathrm{d}^{\,n-1}\br_\parallel\,
C_{\delta \rho F_{\mathrm{int}},z}(\br_\parallel,0^+)
= \kB T\,\rho^2
\int_{\mathcal A} \mathrm{d}^{\,n-1}\br_\parallel\,
\partial_z h(\br_\parallel,0^+).
\end{aligned}
\label{eq:C4}
\end{equation}
Here $\br=(\br_\parallel,z)$, with $\br_\parallel\in\bbR^{\,n-1}$ the in–plane coordinate along the wall and $z$ the normal coordinate; the notation $0^+$ denotes the limit $z\to 0$ taken from the fluid side (inside the accessible region); and $h(\br_\parallel,0^+)$ is the total correlation function $h(\br)=g(\br)-1$ evaluated at the wall, i.e.\ at $\br=(\br_\parallel,z)$ with $z\to0^+$. The identity holds for any finite in-plane domain $\mathcal A$.

\subsubsection{Gauge $\times$ translations: a $k$-derivative identity}
\label{sssec:gauge-trans}

Starting from the master identity in Fourier space, Eq.~\eqref{eq:gauge-master-fourier}, differentiation with respect to $k_\alpha$ ($\alpha,\gamma=1\,.\,.\,n$) produces the translation cross constraint associated with the commutator $[T(\mathbf a),G[\bm\epsilon]]$ in Eq.~\eqref{eq:comm-TG}. Using $\partial_{k_\alpha}(k_\gamma C_{BF_{\mathrm{int}},\gamma})=\delta_{\alpha\gamma}C_{BF_{\mathrm{int}},\gamma}+k_\gamma\partial_{k_\alpha}C_{BF_{\mathrm{int}},\gamma}$ with $\delta_{\alpha\gamma}$ the Kronecker delta, and $\partial_{k_\alpha}k^2=2k_\alpha$, gives
\begin{multline}
\mathrm{i}\,C_{BF_{\mathrm{int}},\alpha}(\bk)
+ \mathrm{i}\,k_\gamma\,\partial_{k_\alpha} C_{BF_{\mathrm{int}},\gamma}(\bk)
= -k_B T\Bigl[
2k_\alpha\,C_{B\rho}(\bk)
\\
+ k^2\,\partial_{k_\alpha} C_{B\rho}(\bk)
\Bigr].
\label{eq:TG-deriv}
\end{multline}

For the density choice $B=\delta\hat\rho(\mathbf 0)$, insertion of Eq.~\eqref{eq:C1} yields
\begin{multline}
\mathrm{i}\,C_{\delta\rho F_{\mathrm{int}},\alpha}(\bk)
+ \mathrm{i}\,k_\gamma\,\partial_{k_\alpha} C_{\delta\rho F_{\mathrm{int}},\gamma}(\bk)
= {} \\
-\,\kB T\Big[
  2k_\alpha\,\rho\,[S(\bk)-1]
  + k^2\,\partial_{k_\alpha}\big(\rho\,[S(\bk)-1]\big)
\Big].
\label{eq:TG-deriv-density}
\end{multline}
Consequently $\mathbf C_{\delta\rho F_{\mathrm{int}}}(\bk)=\mathcal O(k)$ as $k\to0$, consistent with the vanishing of the net internal force at equilibrium.

\subsubsection{Gauge \texorpdfstring{$\times$}{x} momentum--fiber translations}
\label{sssec:gauge-pi}

Define the microscopic current density
\(
\hat{\mathbf j}(\br)=\sum_i \frac{\bp_i}{m_i}\,\delta(\br-\br_i)
\).
For any configuration-only \(B\), the identity with \(\Pi[\bm\chi]\) gives
\(\avg{\cL_{\Pi[\chi]} B}=\beta\avg{B\,\cL_{\Pi[\chi]} H}\).
Since \(\cL_{\Pi[\chi]}B=0\) and \(\cL_{\Pi[\chi]}H=\int \mathrm{d}\br\,\bm\chi(\br)\!\cdot\!\hat{\mathbf j}(\br)\),
arbitrariness of \(\bm\chi\) implies the momentum–fiber/gauge cross constraint
\begin{equation}
\mathbf C_{B j}(\br)\equiv \avg{B\,\hat{\mathbf j}(\br)}=\mathbf 0,
\qquad
\mathbf C_{B j}(\bk)=\mathbf 0.
\label{eq:PiG-current-vanish}
\end{equation}
Thus any equal-time current correlator with a configuration observable vanishes
in equilibrium.  The bracket \([\Pi[\chi],G[\bm\epsilon]]\) in Eq.~\eqref{eq:comm-PiG}
encodes the covariant way this statement is preserved under local gauge shifts.

\subsubsection{Gauge \texorpdfstring{$\times$}{x} boosts (and \texorpdfstring{$t$}{t}-translations):
force--current dynamic identity}
\label{sssec:gauge-boost}

Define the time–domain cross correlators
\[
\mathbf C_{Bj}(\br,t)\equiv \avg{B(0)\,\hat{\mathbf j}(\br,t)},\qquad
C_{B\rho}(\br,t)\equiv \avg{B(0)\,\hat\rho(\br,t)},
\]
and their Fourier transforms
\begin{align}
\mathbf C_{Bj}(\bk,\omega)
&= \int \mathrm{d}^n\br \int_{-\infty}^{\infty} \mathrm{d}t\,
e^{-\mathrm{i}(\bk\cdot\br - \omega t)}\,\mathbf C_{Bj}(\br,t), \\
C_{B\rho}(\bk,\omega)
&= \int \mathrm{d}^n\br \int_{-\infty}^{\infty} \mathrm{d}t\,
e^{-\mathrm{i}(\bk\cdot\br - \omega t)}\,C_{B\rho}(\br,t)
\label{eq:CBj-CBrho-FT-def}
\end{align}

where \(\omega\) is the angular frequency.
Number conservation with time translations
[see Eq.~\eqref{eq:comm-tG}] yields the continuity identity
\begin{equation}
-\,\omega\,C_{B\rho}(\bk,\omega) + \,\bk\!\cdot\!\mathbf C_{Bj}(\bk,\omega)=0.
\label{eq:continuity-ward}
\end{equation}
Combining Eq.~\eqref{eq:continuity-ward} with the gauge master identity,
\(\mathrm{i}\,\bk\!\cdot\!\mathbf C_{BF_{\mathrm{int}}}(\bk,\omega)=-\kB T\,k^2\,C_{B\rho}(\bk,\omega)\),
one obtains the gauge–boost (or gauge–time) cross constraint
\begin{equation}
\mathrm{i}\,\bk\!\cdot\!\mathbf C_{BF_{\mathrm{int}}}(\bk,\omega)
= -\,\kB T\,\frac{k^2}{\omega}\,
\Big[\,\bk\!\cdot\!\mathbf C_{Bj}(\bk,\omega)\Big].
\label{eq:BG-dynamic}
\end{equation}
For single-mass systems, Galilean invariance identifies momentum and current densities, so Eq.~\eqref{eq:BG-dynamic} equivalently links \(\mathbf C_{BF_{\mathrm{int}}}\) and \(\mathbf C_{Bj}\).
In the equal-time limit, both sides vanish consistently with Eq.~\eqref{eq:PiG-current-vanish}.  The appearance of \(\Pi\) in Eq.~\eqref{eq:comm-BG} is the geometric reason why boosts couple the gauge sector to current correlations.

\subsubsection{Gauge $\times$ gauge: integrability and Lie–compatibility}
\label{sssec:gauge-gauge}

The gauge sector closes under commutation, $[G[\bm\epsilon_1],G[\bm\epsilon_2]]=G[[\bm\epsilon_1,\bm\epsilon_2]]$, consistent with the scalar–potential structure of the master identity. It is convenient to write the vector correlator as a differential one–form,
\begin{align}
\varpi_B(\br)\equiv \mathbf C_{BF_{\mathrm{int}}}(\br)\!\cdot\! \exd\br
= \kB T\,\exd C_{B\rho}(\br),
\label{eq:omega-def}
\end{align}
where $\exd$ denotes the exterior differential. Thus $\varpi_B$ is the covariant representation of the same correlator, $\varpi_{B,i}=C_{BF_{\mathrm{int}},i}\,\exd r_i$, which exposes its gradient origin and implies the integrability condition.

\begin{align}
\nabla\times \mathbf C_{BF_{\mathrm{int}}}(\br)=\mathbf 0,
\qquad
\bk\times \mathbf C_{BF_{\mathrm{int}}}(\bk)=\mathbf 0,
\label{eq:Gg-curlfree}
\end{align}
independent of isotropy or central forces.
 In simply connected domains every closed one–form is exact, hence there exists a scalar potential $\Psi_B$ with $\mathbf C_{BF_{\mathrm{int}}}=\nabla\Psi_B$. Equation~\eqref{eq:omega-def} identifies $\Psi_B=\kB T\,C_{B\rho}$, so the longitudinal representation holds globally and circulation vanishes along any closed loop,
\begin{align}
\oint \mathbf C_{BF_{\mathrm{int}}}\!\cdot\! \mathrm{d}\boldsymbol\ell
= \kB T\,\oint \mathrm{d} C_{B\rho}
= 0.
\label{eq:circulation}
\end{align}
In Fourier space a curl–free vector is necessarily longitudinal. Writing $\mathbf C_{BF_{\mathrm{int}}}(\bk)=-\mathrm{i}\,\bk\,\phi_B(\bk)$ and using the divergence identity $\mathrm{i}\,\bk\!\cdot\!\mathbf C_{BF_{\mathrm{int}}}(\bk)=-\kB T\,k^2 C_{B\rho}(\bk)$ gives $\phi_B(\bk)=-\kB T\,C_{B\rho}(\bk)$ and the full vector relation
\begin{align}
\mathbf C_{BF_{\mathrm{int}}}(\bk)=\,\mathrm{i}\,\kB T\,\bk\,C_{B\rho}(\bk),
\label{eq:full-long}
\end{align}
which requires no rotational averaging and refines the longitudinality statement.
Compatibility with the gauge algebra follows from the naturality of the Lie derivative on forms. Using Cartan’s formula
$\cL_{\bm\epsilon}=\exd\circ i_{\bm\epsilon}+i_{\bm\epsilon}\circ \exd$
with $i_{\bm\epsilon}$ the interior contraction with the gauge vector field $\bm\epsilon$, and $\circ$ operator composition, together with Eq.~\eqref{eq:omega-def}, one finds
\begin{align}
\cL_{\bm\epsilon}\,\varpi_B
= \exd\,i_{\bm\epsilon}\varpi_B + i_{\bm\epsilon}\exd\varpi_B
= \kB T\,\exd\big(\cL_{\bm\epsilon} C_{B\rho}\big),
\label{eq:cartan}
\end{align}
so Lie–dragging the one–form equals taking a gradient of the Lie–dragged scalar. The commutator of two such draggings satisfies
\begin{align}
\big(\cL_{\bm\epsilon_1}\cL_{\bm\epsilon_2}-\cL_{\bm\epsilon_2}\cL_{\bm\epsilon_1}\big)\,\varpi_B
= \cL_{[\bm\epsilon_1,\bm\epsilon_2]}\,\varpi_B,
\label{eq:comm-form}
\end{align}
and by Eq.~\eqref{eq:cartan} this reduces to
\begin{align}
\big(\cL_{\bm\epsilon_1}\cL_{\bm\epsilon_2}-\cL_{\bm\epsilon_2}\cL_{\bm\epsilon_1}\big)\,C_{B\rho}
= \cL_{[\bm\epsilon_1,\bm\epsilon_2]}\,C_{B\rho}.
\label{eq:Gg-compat}
\end{align}
The same reasoning applied to $\varpi_B=\kB T\,\exd C_{B\rho}$ shows that the gradient representation is preserved under any finite composition of local gauge flows. In particular, the transverse projector annihilates $\mathbf C_{BF_{\mathrm{int}}}$ in every frame fixed by a gauge diffeomorphism, hence longitudinality does not require $SO(n)$.

The integrability also constrains topology. On periodic domains the potential $C_{B\rho}$ is periodic, hence Eq.~\eqref{eq:circulation} holds and no harmonic one–form component can enter $\varpi_B$. At boundaries Stokes’ theorem converts $\int_{\Sigma} \exd\varpi_B=0$ into a statement about the line integral of $\mathbf C_{BF_{\mathrm{int}}}$ along $\partial\Sigma$, which is the geometric origin of the wall contact relations once the gauge master identity is invoked in the bulk. Altogether, closure of the gauge algebra and exactness of $\varpi_B$ ensure that the scalar potential $C_{B\rho}$ is the unique generator of the mixed force correlator, that the full vector is fixed by its longitudinal projection, and that nested local gauge transformations act compatibly on both scalar and vector sectors.

\subsubsection{Dynamic cross identities and the stress channel}
\label{subsec:dyn-stress}

The dynamic stress relation follows from two geometric inputs. The Irving--Kirkwood identity gives the divergence of the microscopic stress in terms of the internal force density, $\partial_j \hat\sigma_{ij}=\hat F_{{\rm int},i}$, where $\hat\sigma_{ij}$ is the microscopic Irving--Kirkwood stress tensor and $\hat F_{{\rm int},i}$ is the $i$th component of the microscopic internal–force density $\hat{\mathbf F}_{\rm int}$. Hence in Fourier space $\mathrm{i}\,k_j \hat\sigma_{ij}(\bk,t)=\hat F_{{\rm int},i}(\bk,t)$. Taking the mixed correlator with a configuration observable $B$ and contracting with $k_i$ yields
\begin{align}
k_i k_j\,C_{B\sigma_{ij}}(\bk,t)=-\,\mathrm{i}\,\bk\!\cdot\!\mathbf C_{BF_{\mathrm{int}}}(\bk,t),
\label{eq:IK-long}
\end{align}
where $C_{B\sigma_{ij}}(\bk,t)\equiv \int \mathrm{d}^n r\,e^{-\mathrm{i}\bk\cdot\br}\,\big\langle B\,\hat\sigma_{ij}(\br,t)\big\rangle$.

The gauge master identity in the time–frequency domain,
\begin{align}
i\,\bk\!\cdot\!\mathbf C_{BF_{\mathrm{int}}}(\bk,\omega)= -\,\kB T\,k^2\,C_{B\rho}(\bk,\omega),
\label{eq:gauge-omega}
\end{align}
then gives the longitudinal stress constraint
\begin{align}
k_i k_j\,C_{B\sigma_{ij}}(\bk,\omega)=\kB T\,k^2\,C_{B\rho}(\bk,\omega).
\label{eq:dyn-stress-corr}
\end{align}
Equation~\eqref{eq:dyn-stress-corr} holds for any configuration-only $B$ and any state with the same gauge and translational symmetries. Choosing $B=\delta\hat\rho_{-\bk}$ identifies the longitudinal stress kernel by $k_i k_j C_{\delta\rho\sigma_{ij}}\equiv k^2 M_L$, where $M_L(\bk,\omega)$ denotes the longitudinal stress spectrum (kernel) and $C_{\delta\rho\delta\rho}(\bk,\omega)\equiv \int \mathrm{d}^n r\,\mathrm{d}t\,e^{-\mathrm{i}(\bk\cdot\br-\omega t)}\langle \delta\hat\rho(\br,t)\,\delta\hat\rho(\mathbf 0,0)\rangle$ is the dynamic density–density correlator, so that
\begin{align}
M_L(\bk,\omega)=\kB T\,C_{\delta\rho\delta\rho}(\bk,\omega).
\label{eq:ML-equals-density}
\end{align}

The right-hand side is the dynamic density spectrum, proportional to the measured dynamic structure factor under the usual conventions. Nonnegativity of $C_{\delta\rho\delta\rho}(\bk,\omega)$ implies $M_L(\bk,\omega)\ge 0$ at each frequency, and for any nonnegative weight $w(\omega)$ one obtains the sharp integral bound
\begin{align}
\int \mathrm{d}\omega\,w(\omega)\,|M_L(\bk,\omega)|
\le \kB T \int \mathrm{d}\omega\,w(\omega)\,C_{\delta\rho\delta\rho}(\bk,\omega).
\label{eq:bounds}
\end{align}
 Inverse transforming Eq.~\eqref{eq:dyn-stress-corr} gives the equal–time relation $k_i k_j\,C_{B\sigma_{ij}}(\bk,t{=}0)=\kB T\,k^2\,C_{B\rho}(\bk,t{=}0)$; for $B=\delta\hat\rho_{-\bk}$ this reduces to $k_i k_j\,C_{\delta\rho\sigma_{ij}}(\bk,0)=\kB T\,k^2\,[S(\bk)-1]$, which at small $k$ connects the static longitudinal stress to the compressibility via $S(0)=\rho\,\kB T\,\kappa_T$ when $S$ includes the self contribution.

The continuity identity in Eq.~\eqref{eq:continuity-ward}  combines with Eq.~\eqref{eq:dyn-stress-corr} into
\begin{align}
k_i k_j\,C_{B\sigma_{ij}}(\bk,\omega)
= \kB T\,\frac{k^3}{\omega}\,\hat{\bk}\!\cdot\!\mathbf C_{Bj}(\bk,\omega),
\label{eq:stress-current}
\end{align}
which links stress to the longitudinal current channel. In generalized hydrodynamics this identity pins the part of the longitudinal memory function that enters sound and Brillouin peaks. In the hydrodynamic regime one has $C_{\delta\rho\delta\rho}(\bk,\omega)$ controlled by the isothermal sound speed and the longitudinal viscosity; Equation~\eqref{eq:ML-equals-density} therefore fixes the corresponding stress spectrum without direct stress measurements and provides a route from dynamic scattering to the combination $\zeta+4\eta/3$ at low $k$ and small $\omega$, where $\zeta$ is the bulk viscosity and $\eta$ is the shear viscosity.

Causality of the retarded longitudinal stress susceptibility entails analyticity in the upper half-plane and Kramers--Kronig dispersion relations~\cite{klimchitskaya2018kramers}. Through the fluctuation–dissipation theorem~\cite{li2023linear} these analyticity properties are inherited by the correlator representation. Equation~\eqref{eq:ML-equals-density} then implies that the dispersion constraints and sum rules for $M_L$ are the same as those of $C_{\delta\rho\delta\rho}$, including the static sum rule $\int \mathrm{d}\omega\,C_{\delta\rho\delta\rho}(\bk,\omega)=2\pi\,C_{\delta\rho\delta\rho}(\bk,t{=}0)$ and the small–$k$ compressibility limit.

 \subsection{Wigner--Eckart--Ward reduction}
\label{subsec:wew}

Rotational invariance and the gauge master identity combine to fix the tensorial structure of mixed
correlations between an irreducible tensor $T^{(\ell)}$ and the gauge--shift one--body field
$\mathbf J_G$. Here we take a concrete microscopic realization. The rank--$\ell$ irreducible tensor
is built from the local density via spherical harmonics,
\begin{align}
T^{(\ell)}_m(\mathbf 0)
&\equiv \int \mathrm{d}^n r'\, w_\ell(r')\,Y_{\ell m}(\hat{\br'})\,\delta\hat\rho(\br'),
\qquad m=-\ell,\ldots,\ell,
\label{eq:T-def}
\end{align}
with $\hat{\br'}=\br'/r'$ and $w_\ell$ a short--ranged radial weight and $Y_{\ell m}$ the usual harmonics
(bond--orientational choices recover $Q_{\ell m}$~\cite{steinhardt1983bond,lechner2008accurate}).
We take $\mathbf J_G$ as the gauge--shift one--body field~\cite{robitschko2024hyperforce,sammuller2024hyperdensityb}
\begin{align}
\mathbf J_G(\br)\equiv -\,\kB T\,\nabla \hat\rho(\br),
\qquad
\mathbf J_G(\bk)=\,-\mathrm{i}\,\kB T\,\bk\,\hat\rho(\bk),
\label{eq:JG-def}
\end{align}
which ensures $\langle B\,\mathbf J_G(\br)\rangle=-\kB T\,\nabla \langle B\,\hat\rho(\br)\rangle$
for any configuration--only $B$.
We note that $\mathbf J_G$ is different from the observable-dependent hyperforce density of
Schmidt \textit{et al.}~\cite{Matthes2025Gauge}, $\hat{\mathbf S}_A(\br)=\boldsymbol{\sigma}(\br)\hat A$
(with $\boldsymbol{\sigma}$ the local shifting operator, Ref.~\cite{Matthes2025Gauge}, Eq.~(20)).
Here we define $\mathbf J_G$ as the universal one-body gauge (Noether) field entering the
configuration-sector gauge master identity,
$\langle B\,\hat{\mathbf F}_{\rm int}(\br)\rangle=-\langle B\,\mathbf J_G(\br)\rangle$
(up to the standard contact term at the insertion point).
With Eq.~\eqref{eq:JG-def}, $\mathbf J_G(\br)=-k_BT\,\nabla\hat\rho(\br)$, this object is simply the
thermal/entropic (ideal) force density and may be written as the divergence of the ideal stress,
$\mathbf J_G=\nabla\!\cdot\!\hat{\boldsymbol\sigma}^{\rm id}$ with
$\hat{\boldsymbol\sigma}^{\rm id}(\br)= -k_BT\,\hat\rho(\br)\,\mathbf I$.

The symbol $\otimes$ denotes the tensor product of irreducible tensors, so
\begin{align}
\big\langle T^{(\ell)}(\mathbf 0)\otimes \mathbf J_G(\bk)\big\rangle
= \,-\mathrm{i}\,\kB T\,k\,\hat{\bk}\,
\big\langle T^{(\ell)}(\mathbf 0)\,\delta\hat\rho(\bk)\big\rangle,
\label{eq:wew-long}
\end{align}
and all directional content is carried by the angular dependence of the scalar correlator with $\delta\rho$, while the vector character is supplied by $\hat{\bk}$. See App.~\ref{app:wew} for details.

Decompose the scalar spectrum using irreducible harmonics,
\begin{align}
\big\langle T^{(\ell)}(\mathbf 0)\,\delta\hat\rho(\bk)\big\rangle
= \sum_{m=-\ell}^{\ell} \mathcal A_{\ell m}(k)\,Y_{\ell m}(\hat{\bk}),
\label{eq:wew-scalar}
\end{align}
where $\mathcal A_{\ell m}(k)$ are the scalar radial amplitudes obtained by angular projection,
\[
\mathcal A_{\ell m}(k)=\int \frac{\dd \Omega_{\hat{\bk}}}{4\pi}\,
Y_{\ell m}^{*}(\hat{\bk})\,\big\langle T^{(\ell)}(\mathbf 0)\,\delta\hat\rho(\bk)\big\rangle,
\]
which depend only on the wavenumber magnitude $k=|\bk|$ by isotropy; re-expand the product $\hat{\bk}\,Y_{\ell m}$ into irreducible vector or tensor harmonics.
The vector $\hat{\bk}$ transforms as a spherical tensor of rank one, hence the tensor product $1\otimes \ell$ reduces into total angular momentum $L=\ell-1,\ell,\ell+1$. Longitudinality excludes the purely toroidal channel associated with $L=\ell$, since the gradient of a scalar generates only radial and poloidal components. The mixed correlator therefore contains exactly two channels,
\begin{align}
\big\langle T^{(\ell)}(\mathbf 0)\otimes \mathbf J_G(\bk)\big\rangle
&= \sum_{m=-\ell-1}^{\ell+1}\!
\mathcal C_{\ell,1\to \ell+1}\,
\mathcal Y^{(\ell+1)}_{m}(\hat{\bk})\,
\mathcal R_{\ell+1}(k)
\notag\\
&\quad + \sum_{m=-\ell+1}^{\ell-1}\!
\mathcal C_{\ell,1\to \ell-1}\,
\mathcal Y^{(\ell-1)}_{m}(\hat{\bk})\,
\mathcal R_{\ell-1}(k),
\label{eq:WEW}
\end{align}
with direction carried by the irreducible vector or tensor harmonics $\mathcal Y^{(L)}_{m}(\hat{\bk})$ and physics captured by the two scalar radial spectra $\mathcal R_{\ell\pm1}(k)$. The reduced coefficients $\mathcal C_{\ell,1\to L}$ depend only on angular momentum algebra and are independent of $\hat{\bk}$ and $k$.

For $\ell=2$ define $Q_{2m}\equiv T^{(2)}_{m}$ as the bond--orientational components and
$C_{2m,i}(\bk)\equiv\langle Q_{2m}(\mathbf 0)\,J_{G,i}(\bk)\rangle$.
Here $i=-1,0,1$ denotes a spherical vector component (conversion to Cartesian components
is a fixed unitary map and does not affect the radial spectra).
The selection rule gives a decomposition into $L=1$ and $L=3$ only,
\begin{equation}
\begin{aligned}
C_{2m,i}(\bk)
&= -\,\ii\,k\,\kB T
\Bigg[
\alpha_1(k)\sum_{M=-1}^{1}
\langle 2\,m;1\,i|1\,M\rangle\,
Y_{1M}(\hat{\bk})
\\[-2pt]
&\qquad\qquad
+\alpha_3(k)\sum_{M=-3}^{3}
\langle 2\,m;1\,i|3\,M\rangle\,
Y_{3M}(\hat{\bk})
\Bigg] .
\end{aligned}
\label{eq:l2}
\end{equation}

with $\alpha_1$ and $\alpha_3$ the two radial spectra and $\langle 2\,m;1\,i|L\,M\rangle$ Clebsch--Gordan coefficients~\cite{edmonds1996angular}.
The absence of the $L=2$ sector follows from the longitudinal/gradient structure in Eq.~\eqref{eq:wew-long}:
the product $\hat{\bk}\,Y_{2m}(\hat{\bk})$ contains only $L=1$ and $L=3$ (equivalently,
$\langle 2\,0;1\,0|2\,0\rangle=0$), so the toroidal $L=2$ vector--harmonic channel is forbidden.
This expresses the purely longitudinal nature of $\mathbf J_G$, consistent with
$\mathbf P_T(\bk)\,\mathbf C_{\delta\rho F_{\mathrm{int}}}(\bk)=\mathbf 0$.

Orthogonality of the harmonics produces explicit projectors for the radial spectra. Introduce the angular kernels
\begin{align}
\mathcal V^{(1)}_{m,i}(\hat{\bk})
&= \sum_{M=-1}^{1}
\langle 2\,m;1\,i|1\,M\rangle\,
Y_{1M}(\hat{\bk}),
\label{eq:proj-kernels-1} \\
\mathcal V^{(3)}_{m,i}(\hat{\bk})
&= \sum_{M=-3}^{3}
\langle 2\,m;1\,i|3\,M\rangle\,
Y_{3M}(\hat{\bk}).
\label{eq:proj-kernels-2}
\end{align}

and let $\mathcal N_{L}$ denote the corresponding normalization constants fixed by
\begin{equation}
\sum_{m,i}\int \frac{\mathrm{d}\Omega_{\hat{\bk}}}{4\pi}\,
\mathcal V^{(L)}_{m,i}(\hat{\bk})\,\mathcal V^{(L)\,*}_{m,i}(\hat{\bk})
=\mathcal N_L,
\end{equation}
(with $\mathcal N_1=3/(4\pi)$ and $\mathcal N_3=7/(4\pi)$ for the standard convention
$\int \mathrm{d}\Omega\,Y_{LM}^*Y_{L'M'}=\delta_{LL'}\delta_{MM'}$).
The projections read
\begin{align}
\alpha_1(k)
&= \frac{\mathrm{i}}{k\,\kB T}\,
\frac{1}{\mathcal N_1}\,
\sum_{m,i}\!\int\!\frac{\mathrm{d}\Omega_{\hat{\bk}}}{4\pi}\,
\mathcal V^{(1)\,*}_{m,i}(\hat{\bk})\,C_{2m,i}(\bk),
\label{eq:alpha1}\\
\alpha_3(k)
&= \frac{\mathrm{i}}{k\,\kB T}\,
\frac{1}{\mathcal N_3}\,
\sum_{m,i}\!\int\!\frac{\mathrm{d}\Omega_{\hat{\bk}}}{4\pi}\,
\mathcal V^{(3)\,*}_{m,i}(\hat{\bk})\,C_{2m,i}(\bk).
\label{eq:alpha3}
\end{align}
Small wave number scaling follows directly from Eq.~\eqref{eq:l2} (or Eq.~\eqref{eq:wew-long}), which contains an explicit prefactor $k$:
hence $C_{2m,i}(\bk)=\mathcal O(k)$ as $k\to0$ (at fixed reduced amplitudes), and $\alpha_{1}(k)$ and $\alpha_{3}(k)$ remain finite at long wavelength.
The two spectra are therefore genuine radial response functions in the sense of Wigner--Eckart, with all angular dependence stripped away.

The reduction has a direct operational meaning. Angle--resolved or polarized scattering that couples to $Q_{2m}$ isolates the $L=1$ and $L=3$ channels through the kernels in Eqs.~\eqref{eq:proj-kernels-1}--\eqref{eq:proj-kernels-2}, which allows one to extract $\alpha_{1}(k)$ and $\alpha_{3}(k)$ from measured cross responses without reference to stress correlators.
In the time domain the same selection rule applies to longitudinal spectra, once the continuity identity has been folded into the gauge relation, so that dynamic analogues of Eq.~\eqref{eq:l2} follow with $\omega$ dependence carried entirely by the two radial functions.
The Wigner--Eckart--Ward reduction thus provides a compact map from symmetry to measurable amplitudes, explains the vanishing of transverse components, and collapses tensor--hyperforce correlations to two scalar spectra in isotropic central--force fluids.

\subsection{Equivariant gauge-constrained DFT}
\label{subsec:eqdft}

The Ward and cross--Ward identities link structure and mechanics at equilibrium. To build the gauge constraint already at the level of stationary points, we extend the usual DFT field content to the multiplet
$\Phi=(\rho,\mathbf J_G)$, where $\rho(\br)$ is the one--body density and $\mathbf J_G(\br)$ is the gauge--shift one--body field. 
The role of the auxiliary one-body field $\mathbf J_G$ is to represent the universal ideal/entropic
force density that appears on the right-hand side of the configuration-sector gauge master relation.
Using the microscopic definition already introduced in Eq.~\eqref{eq:JG-def}, one has for any configuration observable $B(\br^N)$
\begin{equation}
\avg{B\,\mathbf J_G(\br)}
=-\,\kB T\,\nabla \avg{B\,\hat\rho(\br)},
\label{eq:JG-density-link}
\end{equation}
Inserting Eq.~\eqref{eq:JG-density-link} into the configuration-sector hyperforce identity
Eq.~\eqref{eq:connection-hyperYBG} yields the equivalent force-balance form
\begin{equation}
\begin{aligned}
\avg{B\,\hat{\mathbf F}_{\rm int}}
-\avg{B\,\hat\rho}\,\nabla V_{\rm ext}(\br)
&=
-\,\avg{B\,\mathbf J_G}
\\
&\quad
-\kB T
\Big\langle
\sum_i \delta(\br-\br_i)\,\nabla_{\br_i}B
\Big\rangle .
\end{aligned}
\label{eq:JG-Fint-link}
\end{equation}
For the translation-covariant insertions used throughout the bulk discussion (e.g.\
$B=\delta\hat\rho(\mathbf 0)$), the last term in Eq.~\eqref{eq:JG-Fint-link} is a local contact contribution at the
insertion point and vanishes for $\br\neq\mathbf 0$. Thus, away from the contact term one recovers the compact bulk
master relation
$\avg{B\,\hat{\mathbf F}_{\rm int}(\br)}=-\avg{B\,\mathbf J_G(\br)}$,
which is equivalent to the gauge--$\mathrm U(1)$ master identity
Eqs.~\eqref{eq:gauge-master-real}--\eqref{eq:gauge-master-fourier}.

The hyperdensity functional theory of Ref.~\cite{sammuller2024hyperdensityb}
constructs density functionals for general observables $A[\rho]$ via the hyperdirect
correlation functional $c_A(\br,[\rho])=\delta A[\rho]/\delta\rho(\br)$ together with a
hyper--Ornstein--Zernike relation for the hyperfluctuation profile $\chi_A$.
Our construction here has a different purpose: we keep the physical Hamiltonian fixed and
introduce $\mathbf J_G$ as an auxiliary field so that the stationary Euler--Lagrange
equations enforce the gauge Ward identity and its correlated consequences.

In standard classical DFT the intrinsic free energy is decomposed as
$\mathscr F[\rho]=\mathscr F_{\rm id}[\rho]+\mathscr F_{\rm ex}[\rho]$.
The ideal-gas contribution is given explicitly by
\begin{equation*}
\mathscr F_{\rm id}[\rho]
=\kB T\int \dd^n\br\,\rho(\br)\Big[\ln\!\big(\rho(\br)\Lambda^n\big)-1\Big],
\end{equation*}
where $\Lambda$ is the constant thermal de Broglie wavelength; the additive constant $\ln\Lambda^n$ can be absorbed into a shift of $\mu$.
We take the excess part as
$\mathscr F_{\rm ex}[\rho]=\mathscr F_{\rm HS}^{\rm FMT}[\rho]+\mathscr F_{\rm tail}^{\rm MF}[\rho]$, where $\mathscr F_{\rm ex}[\rho]$ is the interaction part of the intrinsic Helmholtz free-energy functional. We approximate it by a hard-sphere reference treated in fundamental measure theory (FMT), $\mathscr F_{\rm HS}^{\rm FMT}[\rho]$, plus a mean-field (MF) contribution from the non-hard-sphere (typically attractive) tail,
\begin{equation}
\mathscr F_{\rm tail}^{\rm MF}[\rho]
=\frac{1}{2}\int \dd^n\br\int \dd^n\br'\,
\rho(\br)\rho(\br')\,w_{\rm tail}(|\br-\br'|),
\label{eq:FtailMF}
\end{equation}
with $w_{\rm tail}(r)$ the chosen tail part of the pair interaction.

We work with the grand potential functional
\begin{align}
\Omega[\rho,\mathbf J_G]
&=
\mathscr F_{\rm id}[\rho]
+\mathscr F_{\rm HS}^{\rm FMT}[\rho]
+\mathscr F_{\rm tail}^{\rm MF}[\rho]
\notag\\
&\quad
+\int \dd^n\br\,\rho(\br)\big[V_{\rm ext}(\br)-\mu\big]
\notag\\
&\quad
+\frac{1}{2\lambda_G}
\int \dd^n\br\,
\big|\mathbf J_G(\br)+\kB T\,\nabla\rho(\br)\big|^2 .
\label{eq:minF}
\end{align}

The first line gives the intrinsic free energy (ideal, FMT hard-sphere, and mean-field tail), the second line couples to $V_{\rm ext}$ and $\mu$, and the third line imposes the gauge constraint through the positive penalty modulus $\lambda_G$.
In bulk ($V_{\rm ext}=0$) the baseline DFT functional $\Omega_0[\rho]=\mathscr F_{\rm id}[\rho]+\mathscr F_{\rm HS}^{\rm FMT}[\rho]+\mathscr F_{\rm tail}^{\rm MF}[\rho] +\int \dd^n\br\,\rho(\br)\big[V_{\rm ext}(\br)-\mu\big]$ is already Euclidean invariant by construction.
The role of the penalty term is instead to impose the gauge Ward constraint, which controls mixed force--density (and associated stress) correlations and is not automatically enforced by a practical density-only approximation.
Introducing the auxiliary field $\mathbf J_G$ via the covariant combination $\mathbf J_G+\kB T\nabla\rho$ enforces the gauge master identity at the Euler--Lagrange level.
In bulk one has $V_{\rm ext}=0$. For hard confinement, $V_{\rm ext}$ is implemented by restricting the integration domain to the accessible region and keeping the associated surface term generated by integration by parts, as in Eq.~\eqref{eq:boundary} below.

Stationarity of $\Omega$ yields
\begin{align}
\frac{\delta \Omega}{\delta \mathbf J_G}
&=\frac{1}{\lambda_G}\big(\mathbf J_G+\kB T\,\nabla \rho\big)
=\mathbf 0
\qquad \quad
\mathbf J_G=-\kB T\,\nabla\rho,
\label{eq:ELJ}\\[0.5ex]
\frac{\delta \Omega}{\delta \rho}
&=
\frac{\delta \mathscr F_{\rm id}}{\delta \rho}
+\frac{\delta \mathscr F_{\rm HS}^{\rm FMT}}{\delta \rho}
+\frac{\delta \mathscr F_{\rm tail}^{\rm MF}}{\delta \rho}
+V_{\rm ext}(\br)
\nonumber\\
&\quad
-\frac{\kB T}{\lambda_G}\,
\diver\big(\mathbf J_G+\kB T\,\nabla\rho\big)
=\mu.
\label{eq:ELrho}
\end{align}

Inserting Eq.~\eqref{eq:ELJ} into Eq.~\eqref{eq:ELrho} removes the penalty divergence and recovers the standard DFT Euler equation
$\delta \mathscr F_{\rm id}/\delta\rho+\delta \mathscr F_{\rm HS}^{\rm FMT}/\delta \rho
+\delta \mathscr F_{\rm tail}^{\rm MF}/\delta \rho+V_{\rm ext}=\mu$.

The external field selects which geometric symmetries remain unbroken. Accordingly, the variational Ward form applies to the generators $X$ of the unbroken subgroup, for which $\Omega$ is equivariant:
\begin{equation}
\sum_i \int \dd^n\br\,
\frac{\delta \Omega}{\delta \Phi_i}\,
(\cL_X \Phi_i)=0,
\label{eq:master-ward}
\end{equation}
for all $X\in\mathrm{Lie}(\cG)$ unbroken by $V_{\rm ext}$.

Upon averaging, this becomes
\begin{align}
\sum_i \big\langle \Lambda_i\,\cL_X \Phi_i \big\rangle=0,
\qquad
\Lambda_i\equiv\frac{\delta\Omega}{\delta\Phi_i},
\label{eq:eqDFT}
\end{align}
which matches the thermal Ward form derived from Liouville--preserving flows.

The field--level constraint propagates to correlations. In bulk, linearizing $\Omega$ around a stationary state and Fourier transforming gives a quadratic form whose completion of the square eliminates $\delta\mathbf J_{G,\bk}$ and yields
\begin{align}
\delta\mathbf J_{G,\bk}^{\ \rm stat}=\,-\mathrm{i}\,\kB T\,\bk\,\delta\hat\rho_{\bk}.
\label{eq:gauss}
\end{align}
For any configuration--only probe $B$ this implies
\begin{align}
\mathrm{i}\,\bk\!\cdot\!\avg{B\,\delta\mathbf J_{G,\bk}}
= \,\kB T\,k^2\,\avg{B\,\delta\hat\rho_{\bk}},
\label{eq:Fourier-ward-functional}
\end{align}
which is the equal--time Fourier form of the gauge master identity in the bulk and coincides with Eq.~\eqref{eq:gauge-master-fourier}.
In inhomogeneous settings one keeps the corresponding real--space form
$\avg{B\,\delta\mathbf J_G(\br)}=-\kB T\,\nabla\avg{B\,\delta\rho(\br)}$ rather than plane waves.

Combining Eq.~\eqref{eq:Fourier-ward-functional} with the Irving--Kirkwood relation produces the longitudinal stress constraint
\begin{align}
k_i k_j\,C_{B\sigma_{ij}}(\bk)=\kB T\,k^2\,C_{B\rho}(\bk),
\label{eq:stress-from-density}
\end{align}
which underpins the scattering--to--stress route used later.

Surface contributions arise from the penalty term by integration by parts. Introducing an arbitrary smooth test field $\mathbf q(\br)$, one writes
\begin{equation}
\begin{aligned}
\int \mathrm{d}^n\br\,\mathbf q\!\cdot\!\kB T\,\nabla\rho
&=
-\,\kB T \int \mathrm{d}^n\br\,(\nabla\!\cdot\!\mathbf q)\,\rho
\\
&\quad
+\,\kB T \oint_{\partial\Omega}\! \mathrm{d}S\,
(\mathbf q\!\cdot\!\mathbf n)\,\rho .
\end{aligned}
\label{eq:boundary}
\end{equation}

Periodic domains have no boundary term, whereas walls contribute a normal contact piece, consistent with Eq.~\eqref{eq:C4} once the bulk Ward identity is invoked.

The auxiliary-field formulation is exactly equivalent to standard DFT at the level of the
stationary one--body profile: eliminating $\mathbf J_G$ through Eq.~\eqref{eq:ELJ} reduces
Eq.~\eqref{eq:minF} to $\Omega_0[\rho]$ and thus leaves the Euler equation for $\rho$ unchanged.
The utility of introducing $\mathbf J_G$ is instead in the fluctuation/response sector:
the quadratic expansion of $\Omega[\rho,\mathbf J_G]$ around a stationary state generates a coupled
Gaussian kernel for $(\delta\rho,\delta\mathbf J_G)$, whose saddle relation in Eq.~\eqref{eq:gauss}
yields the correlator constraint
Eq.~\eqref{eq:Fourier-ward-functional}. This is the variational entry point for the stress and
cross--Ward relations in Sec.~\ref{subsec:dyn-stress}. The parameter $\lambda_G$ controls the cost of
fluctuations transverse to the constraint (and can improve numerical conditioning), while the
Ward relation for mixed $\rho$--$\mathbf J_G$ correlators remains fixed by the covariant structure
$\mathbf J_G+\kB T\nabla\rho$.

Finally, for observables $B(\br^N,\bp^N)$ with explicit momentum dependence, the microscopic gauge Ward identity contains additional kinematic source terms generated by the momentum part of the gauge generator. A field-level implementation of those identities would require extending the multiplet $\Phi$ beyond $(\rho,\mathbf J_G)$, for example by adding the current or kinetic-stress sector. The present minimal functional is therefore tailored to the configuration sector used throughout this work.

\section{Simulation}

\begin{figure}
  \centering
  \includegraphics[width=\linewidth]{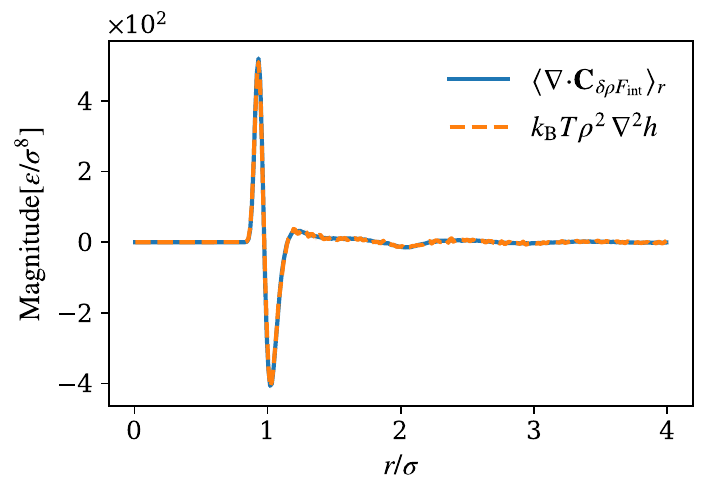}
  \caption{Radial validation of the gauge–\(\mathrm U(1)\) identity \(\diver \mathbf C_{\delta\rho F_{\mathrm{int}}}=\kB T\,\rho^2\,\nabla^2 h\) in a bulk Lennard–Jones fluid (\(T^\ast=1.5\), \(\rho^\ast=0.85\)).
  Solid line: shell–averaged divergence \(\langle \diver \mathbf C_{\delta\rho F_{\mathrm{int}}}\rangle_r\) computed from the density–internal–force cross–correlator measured in MD.
  Dashed line: \(\kB T\,\rho^2\,\nabla^2 h(r)\) obtained from the total correlation \(h(r)=g(r)-1\) of the same run.
  }
  \label{fig:1}
\end{figure}

We simulate a one--component Lennard--Jones (LJ) fluid in a cubic periodic box containing \(N = 6912\) particles. The pair potential is
\[
u(r) = 4\varepsilon \!\left[ \left( \frac{\sigma}{r} \right)^{12} - \left( \frac{\sigma}{r} \right)^6 \right],
\]
where \(\varepsilon\) is the characteristic energy scale corresponding to the depth of the potential well, while \(\sigma\) specifies the characteristic length scale that defines the effective particle diameter. The potential is truncated at \(r_c = 2.5\sigma\) without applying long--range tail corrections.

Molecular dynamics trajectories are generated with \textsc{HOOMD-blue}~\cite{anderson2020hoomd,glaser2015strong} using a velocity--Verlet integrator. Temperature is controlled with a Bussi stochastic velocity--rescaling thermostat (canonical sampling), and the simulations are performed at reduced temperature \(T^{\ast}=k_{\mathrm B}T/\varepsilon\) and reduced density \(\rho^{\ast}=\rho\sigma^{3}\). Each system is equilibrated for \(10^{7}\) integration steps followed by a production run of \(10^{6}\) steps.

Figure~\ref{fig:1} shows stringent real-space validation of the gauge--$\mathrm U(1)$
divergence identity in Eq.~\eqref{eq:C1} for the bulk Lennard--Jones state point
$(T^\ast=1.5,\rho^\ast=0.85)$.
The two sides of Eq.~\eqref{eq:C1}--the shell-averaged divergence
$\langle \nabla\!\cdot\!\mathbf C_{\delta\rho F_{\mathrm{int}}}\rangle_r$
measured directly from the trajectory and the structural expression
$\kB T\,\rho^2\,\nabla^2 h(r)$ obtained independently from $h(r)=g(r)-1$--are essentially
indistinguishable on the scale of the plot over the full $r$ range shown.
In particular, the strong near-contact peak, the subsequent negative lobe at the first minimum,
and the damped oscillations across successive coordination shells coincide in both phase and
amplitude, demonstrating that the mixed density--internal-force response is completely fixed by
the equilibrium pair structure as predicted by the Ward identity.
At larger separations both signals rapidly decay toward zero and remain consistent with zero
within the remaining statistical noise, as expected for a homogeneous bulk fluid with short-ranged
interactions.

\begin{figure}
  \centering
  \includegraphics[width=\linewidth]{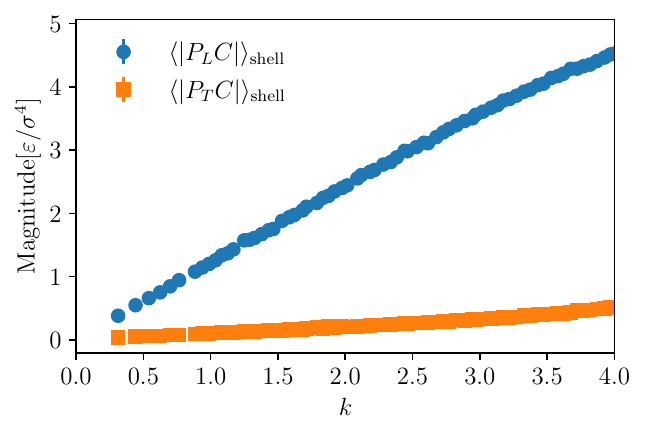}
  \caption{Shell‑averaged magnitudes of the longitudinal (circles) and transverse (squares) projections of the density–internal‑force cross correlator
\(\mathbf C_{\delta\rho F_{\mathrm{int}}}(\bk)
= V^{-1}\big\langle \delta\hat\rho_{-\bk}\,\hat{\mathbf F}_{\mathrm{int}}(\bk)\big\rangle
\)
for a bulk Lennard–Jones fluid at \(T^\ast=1.5\) and \(\rho^\ast=0.85\). The transverse signal is strongly suppressed relative to the longitudinal one,
consistent with the gauge\(\times\)rotation cross identity
\(\mathbf P_T(\bk)\,\mathbf C_{\delta\rho F_{\mathrm{int}}}(\bk)=\mathbf 0.
\)
[Eq.~\eqref{eq:C2}], while the small‑\(k\) growth of the longitudinal branch
follows the gauge–\(\mathrm U(1)\) prediction
\(|P_L\mathbf C_{\delta\rho F_{\mathrm{int}}}(\bk)| \propto k\,|S(k)-1|
\) from Eq.~\eqref{eq:C1}.
}
  \label{fig:2}
\end{figure}

In Fig.~\ref{fig:2} we resolve the Fourier–space structure of the density–internal–force cross correlator,
\(\mathbf C_{\delta\rho F_{\mathrm{int}}}(\bk)=V^{-1}\avg{\delta\hat\rho_{-\bk}\,\mathbf F_{\mathrm{int}}(\bk)}\),
by projecting onto the longitudinal and transverse sectors with
\(P_L(\bk)=\hat\bk\hat\bk^\top\) and \(P_T(\bk)=\Id-P_L(\bk)\), followed by thin \(|\bk|\)-shell averaging of the magnitudes.
Across all accessible wave numbers the signal is overwhelmingly longitudinal:
\(\langle |P_T\,\mathbf C_{\delta\rho F_{\mathrm{int}}}|\rangle_{\rm shell}
\ll \langle |P_L\,\mathbf C_{\delta\rho F_{\mathrm{int}}}|\rangle_{\rm shell}\).
This pure longitudinality is the expected consequence of rotational invariance combined with the gauge identity, Eq.~\eqref{eq:C2}, which forbids a transverse component in an isotropic, central-force fluid. The small but finite transverse baseline is attributable to finite statistics, discrete anisotropy of the reciprocal grid, and the use of shell averages of absolute values. 

The longitudinal amplitude grows approximately linearly with \(k\) at small wave number, in agreement with the gauge–\(\mathrm U(1)\) master identity, Eq.~\eqref{eq:C1}, which gives
\(P_L\,\mathbf C_{\delta\rho F_{\mathrm{int}}}(\bk)=\,\mathrm{i}\,\hat\bk\,k\,\kB T\,\rho\,[S(k)-1]\)
and therefore \(|P_L\,\mathbf C_{\delta\rho F_{\mathrm{int}}}|\propto k\,|S(k)-1|\) as \(k\!\to\!0\).
Figure~\ref{fig:2} thus provides Fourier-space evidence for the gauge\(\times\)rotation cross-Ward constraint and illustrates the reduction of the vector hyperforce response to a single radial spectrum in isotropic liquids.

The isothermal compressibility \(\kappa_T\) was extracted by two independent routes. Using the compressibility relation
\(\kappa_T=S(0)/(\rho\,\kB T)\) with \(S(0)\) evaluated including the self
contribution, we obtain the value
\(\kappa_T=3.612\times10^{-2}\). 
An independent estimate follows from the gauge–\(\mathrm U(1)\) master identity
[Eq.~\eqref{eq:C1}] gives \(\kappa_T=3.595(70)\times10^{-2}\). The calculated isothermal compressibility values from both methods are in good agreement.

The two estimators provide a parameter-free check of the first-moment (gauge$\times$dilation) identity, Eq.~\eqref{eq:C3a}; the resulting values $3.664\times10^{-2}$ and $3.648\times10^{-2}$ are in close agreement, confirming that the same $\kappa_T$ governs both the density--density and density--hyperforce sectors.

\begin{figure}[t]
  \centering
  \includegraphics[width=\linewidth]{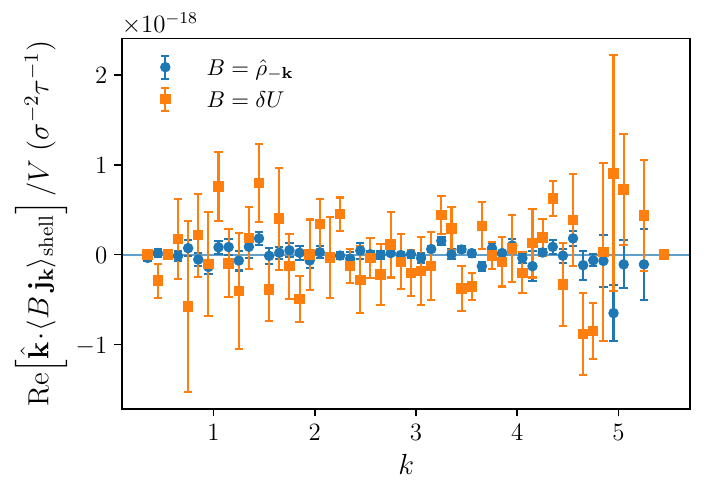}
\caption{
Test of the momentum--fiber/gauge identity, Eq.~\eqref{eq:PiG-current-vanish}, using the signed longitudinal current cross correlator
$\mathrm{Re}\!\left[\hat{\bk}\!\cdot\!\langle B\,\hat{\mathbf j}_{\bk}\rangle_{\rm shell}\right]/V$
(circles: $B=\hat\rho_{-\bk}$; squares: $B=\delta U$).
The exact prediction is zero (horizontal line); error bars are block-averaged standard errors.
Units are Lennard--Jones reduced, with $\tau\equiv\sigma\sqrt{m/\varepsilon}$.
}
\label{fig:rhojUj}
\end{figure}
We test the momentum--fiber/gauge cross identity, Eq.~\eqref{eq:PiG-current-vanish},
which predicts vanishing equal-time current cross correlators with any configuration observable $B$.
We evaluate instead the signed longitudinal shell average
$\mathrm{Re}\!\left[\hat{\bk}\!\cdot\!\langle B\,\hat{\mathbf j}_{\bk}\rangle_{\rm shell}\right]/V$
for two representative choices: $B=\hat\rho_{-\bk}$ and $B=\delta U$. As shown in Fig.~\ref{fig:rhojUj}, both channels fluctuate about zero with no systematic drift over the
accessible $k$ range, and remain consistent with the exact prediction within block-averaged standard errors.
The $\delta U$ channel exhibits larger scatter, reflecting the larger variance of the global potential-energy
observable and correspondingly slower statistical convergence, but is likewise compatible with zero.

\section{Conclusion}
\label{sec:conclusion}

We have formulated a single symmetry framework for equilibrium many‑body systems by embedding standard geometric and dynamical invariances together with the phase‑space gauge‑shifting symmetry into the unified group $\cG$. Via Noether’s calculus for Liouville‑preserving flows we derived a hierarchy of Ward identities and, crucially, new cross‑Ward constraints that originate from the noncommutation of generators. The gauge–$\mathrm U(1)$ master identity fixes force–density cross correlators [Eqs.~\eqref{eq:gauge-master-real}–\eqref{eq:C1}]; in concert with rotations it enforces strict longitudinality and curl‑freeness [Eqs.~\eqref{eq:C2}, \eqref{eq:Gg-curlfree}]; with dilations it yields a first‑moment/compressibility sum rule [Eqs.~\eqref{eq:C3a}, \eqref{eq:C3b}]; with translations it implies a $k$‑derivative identity [Eq.~\eqref{eq:TG-deriv-density}]; with boundaries it produces a wall‑contact relation [Eq.~\eqref{eq:C4}]; and with boosts/time translations it links force–, current–, and stress–response channels, fixing the longitudinal stress from density dynamics and providing bounds [Eqs.~\eqref{eq:BG-dynamic}–\eqref{eq:bounds}]. The Wigner–Eckart–Ward reduction further collapses tensor–hyperforce correlators to a small set of scalar radial spectra in isotropic fluids [Eqs.~\eqref{eq:WEW}, \eqref{eq:l2}]. We also outlined an equivariant gauge-constrained DFT whose Euler–Lagrange equations automatically satisfy the full set of identities [Eqs.~\eqref{eq:minF}–\eqref{eq:eqDFT}].

The unified gauge–geometry perspective offers a systematic organizer for structure–mechanics relations in liquids, mixtures, and interfaces, and suggests practical routes to infer elastic/viscoelastic kernels from static or angle‑resolved scattering. Extensions to multicomponent electrolytes~\cite{petersen2024toward}, confined~\cite{Sermoud2024cDFTConfined,Sammueller2023PRE} or anisotropic media~\cite{Simon2025JCP} and quantum statistics~\cite{li2023linear,Muller2025QuantumGauge} are natural next steps.

\appendix
\section{Derivation of the nonzero commutators}
\label{app:commutators}

This Appendix derives Eqs.~\eqref{eq:comm-TG}--\eqref{eq:comm-tG} by direct evaluation of Lie brackets of the
phase-space vector fields that represent the generators.
Throughout we work at fixed particle number \(N\) and write the phase-space point
\(\Gamma=(\br^N;\bp^N)\).
All generators below are first-order differential operators acting on smooth
phase-space functions \(F(\Gamma)\); for such operators the commutator
\([U,V]\equiv UV-VU\) coincides with the Lie bracket of the associated vector fields.

\subsection{Conventions and one-particle reduction}
For each particle \(i\) and Cartesian component \(\alpha=1,\dots,n\) we use
\[
\partial_{i\alpha}\equiv \frac{\partial}{\partial r_{i\alpha}},
\qquad
\bar\partial_{i\alpha}\equiv \frac{\partial}{\partial p_{i\alpha}},
\qquad
\nabla_i\equiv (\partial_{i1},\dots,\partial_{in}).
\]
For a phase-space vector field
\[
U=\sum_{i=1}^N\Big(U^{r}_{i\alpha}\,\partial_{i\alpha}+U^{p}_{i\alpha}\,\bar\partial_{i\alpha}\Big),
\]
the Lie bracket is
\begin{equation}
[U,V]= (U\!\cdot\!\nabla_\Gamma)\,V-(V\!\cdot\!\nabla_\Gamma)\,U,
\label{eq:app-liebracket}
\end{equation}
equivalently \([U,V]F=U(VF)-V(UF)\) for any test function \(F(\Gamma)\).
Here \(\nabla_\Gamma\) denotes the full gradient with respect to all
\((\br_i,\bp_i)\).

All generators used in Sec.~\ref{subsec:lie} are sums of one-particle pieces,
\(U=\sum_i U_i\), where \(U_i\) acts only on \((\br_i,\bp_i)\).
Hence \([U_i,V_j]=0\) for \(i\neq j\), and therefore
\begin{equation}
[U,V]=\sum_{i=1}^N [U_i,V_i].
\label{eq:app-sumreduce}
\end{equation}
It is thus sufficient to compute each bracket for a single particle and then sum over \(i\).

\subsection{Generators in components}
We use the phase-space representatives (Sec.~\ref{subsec:lie}):
\begin{align}
T(\mathbf a) &= \sum_i a_\alpha\,\partial_{i\alpha}, \label{eq:app-T}\\
R(\Theta) &= \sum_i\Big[(\Theta\br_i)_\alpha\,\partial_{i\alpha}+(\Theta\bp_i)_\alpha\,\bar\partial_{i\alpha}\Big], \label{eq:app-R}\\
D &= \sum_i\Big[r_{i\alpha}\,\partial_{i\alpha}-p_{i\alpha}\,\bar\partial_{i\alpha}\Big], \label{eq:app-D}\\
\mathcal B(\mathbf u) &= \sum_i\Big[t\,u_\alpha\,\partial_{i\alpha}+m_i u_\alpha\,\bar\partial_{i\alpha}\Big], \label{eq:app-B}\\
\Pi[\bm\chi] &= \sum_i \chi_\alpha(\br_i)\,\bar\partial_{i\alpha}, \label{eq:app-Pi}
\end{align}
and the gauge-shift (cotangent-lift) generator
\begin{equation}
G[\bm\epsilon]=\sum_{i=1}^N G_i[\bm\epsilon],
\qquad
G_i[\bm\epsilon]
=\bm\epsilon_\alpha(\br_i)\,\partial_{i\alpha}
-\big(\partial_{i\beta}\bm\epsilon_\alpha(\br_i)\,p_{i\beta}\big)\,\bar\partial_{i\alpha},
\label{eq:app-G}
\end{equation}
where \((\nabla\bm\epsilon)_{\alpha\beta}=\partial_\beta\bm\epsilon_\alpha\).
When needed we also use the spatial Lie bracket of vector fields,
\begin{equation}
[\bm\epsilon_1,\bm\epsilon_2]_\alpha
=\bm\epsilon_{1,\beta}\partial_\beta \bm\epsilon_{2,\alpha}
-\bm\epsilon_{2,\beta}\partial_\beta \bm\epsilon_{1,\alpha}.
\label{eq:app-spatialbracket}
\end{equation}

\subsection{Translation--gauge commutator}
Let \(T_i(\mathbf a)=a_\alpha\partial_{i\alpha}\) and \(G_i[\bm\epsilon]\) as in \eqref{eq:app-G}.
Using \eqref{eq:app-liebracket} and noting that \(T_i\) has constant coefficients,
the \(r\)-component of \([T_i,G_i]\) is
\[
([T_i,G_i])^r_\alpha
=T_i\big(\bm\epsilon_\alpha(\br_i)\big)-G_i(a_\alpha)
=a_\beta \partial_{i\beta}\bm\epsilon_\alpha(\br_i).
\]
The \(p\)-component is obtained by acting \(T_i\) on
\(-\partial_{i\beta}\bm\epsilon_\alpha(\br_i)\,p_{i\beta}\):
\[
([T_i,G_i])^p_\alpha
=T_i\!\Big(-\partial_{i\beta}\bm\epsilon_\alpha(\br_i)\,p_{i\beta}\Big)
=-\partial_{i\beta}\big(a_\gamma\partial_{i\gamma}\bm\epsilon_\alpha(\br_i)\big)\,p_{i\beta}.
\]
Therefore \([T_i,G_i]=G_i[(\mathbf a\!\cdot\!\nabla)\bm\epsilon]\), and summing over \(i\) gives
\begin{equation}
[T(\mathbf a),G[\bm\epsilon]]=G\!\big[(\mathbf a\!\cdot\!\nabla)\bm\epsilon\big].
\label{eq:app-comm-TG}
\end{equation}

\subsection{Rotation--gauge commutator and the component form of Eq.~(9)}
For one particle,
\(
R_i(\Theta)=(\Theta\br_i)_\alpha\partial_{i\alpha}+(\Theta\bp_i)_\alpha\bar\partial_{i\alpha}
\)
and \(G_i[\bm\epsilon]\) as in \eqref{eq:app-G}.
First, the \(r\)-component:
\begin{align*}
([R_i,G_i])^r_\alpha
&=R_i\!\big(\bm\epsilon_\alpha(\br_i)\big)-G_i\!\big((\Theta\br_i)_\alpha\big) \\
&=(\Theta\br_i)_\beta\partial_{i\beta}\bm\epsilon_\alpha(\br_i)
-\bm\epsilon_\beta(\br_i)\,\partial_{i\beta}\big((\Theta\br_i)_\alpha\big) \\
&=(\Theta\br_i)_\beta\partial_{i\beta}\bm\epsilon_\alpha(\br_i)-\bm\epsilon_\beta(\br_i)\,\Theta_{\alpha\beta}.
\end{align*}
Thus the rotated parameter field is
\begin{equation}
\big[(\Theta\br\!\cdot\!\nabla)\bm\epsilon-\Theta\bm\epsilon\big]_\alpha
=(\Theta\br)_\beta\,\partial_\beta \bm\epsilon_\alpha-\Theta_{\alpha\beta}\bm\epsilon_\beta,
\label{eq:app-rotparam-components}
\end{equation}
which is manifestly a spatial vector field and therefore a valid argument of \(G[\cdot]\).

Next, the \(p\)-component.
Write \(G_i^p{}_\alpha=-\partial_{i\beta}\bm\epsilon_\alpha\,p_{i\beta}\).
Then
\begin{equation}
\begin{aligned}
([R_i,G_i])^p_\alpha
={}&(\Theta\br_i)_\gamma
\,\partial_{i\gamma}
\!\Big(-\partial_{i\beta}\bm\epsilon_\alpha\,p_{i\beta}\Big)
\\
&+(\Theta\bp_i)_\gamma
\,\bar\partial_{i\gamma}
\!\Big(-\partial_{i\beta}\bm\epsilon_\alpha\,p_{i\beta}\Big)
\\
&-\Big(-\partial_{i\beta}\bm\epsilon_\gamma\,p_{i\beta}\Big)
\,\bar\partial_{i\gamma}
\big((\Theta\bp_i)_\alpha\big).
\end{aligned}
\end{equation}

Since \(\bar\partial_{i\gamma}(p_{i\beta})=\delta_{\gamma\beta}\) and
\(\bar\partial_{i\gamma}((\Theta\bp_i)_\alpha)=\Theta_{\alpha\gamma}\), this becomes
\begin{equation}
\begin{aligned}
([R_i,G_i])^p_\alpha
={}&-(\Theta\br_i)_\gamma
\,(\partial_{i\gamma}\partial_{i\beta}\bm\epsilon_\alpha)\,p_{i\beta}
\\
&-(\Theta\bp_i)_\beta
\,\partial_{i\beta}\bm\epsilon_\alpha
\\
&+\Theta_{\alpha\gamma}
\,(\partial_{i\beta}\bm\epsilon_\gamma)\,p_{i\beta}.
\end{aligned}
\end{equation}

Using \((\Theta\bp_i)_\beta=\Theta_{\beta\gamma}p_{i\gamma}\) one checks that this equals
\[
([R_i,G_i])^p_\alpha
=-\partial_{i\beta}\Big((\Theta\br_i)_\gamma\partial_{i\gamma}\bm\epsilon_\alpha-\Theta_{\alpha\gamma}\bm\epsilon_\gamma\Big)\,p_{i\beta},
\]
i.e. the cotangent-lift form with parameter \((\Theta\br\!\cdot\!\nabla)\bm\epsilon-\Theta\bm\epsilon\).
Summing over \(i\) yields
\begin{equation}
[R(\Theta),G[\bm\epsilon]]
=G\!\big[(\Theta\br\!\cdot\!\nabla)\bm\epsilon-\Theta\bm\epsilon\big].
\label{eq:app-comm-RG}
\end{equation}

\subsection{Dilation--gauge commutator}
For one particle \(D_i=r_{i\alpha}\partial_{i\alpha}-p_{i\alpha}\bar\partial_{i\alpha}\).
The \(r\)-component is
\begin{align*}
([D_i,G_i])^r_\alpha
&=D_i\big(\bm\epsilon_\alpha(\br_i)\big)-G_i(r_{i\alpha}) \\
&=r_{i\beta}\partial_{i\beta}\bm\epsilon_\alpha(\br_i)-\bm\epsilon_\alpha(\br_i)
=\big[(\br\!\cdot\!\nabla)\bm\epsilon-\bm\epsilon\big]_\alpha.
\end{align*}
The \(p\)-component follows similarly and collapses to the cotangent-lift form
\[
([D_i,G_i])^p_\alpha
=-\partial_{i\beta}\Big((r_{i\gamma}\partial_{i\gamma}\bm\epsilon_\alpha)-\bm\epsilon_\alpha\Big)p_{i\beta}.
\]
Hence
\begin{equation}
[D,G[\bm\epsilon]]=G\!\big[(\br\!\cdot\!\nabla)\bm\epsilon-\bm\epsilon\big].
\label{eq:app-comm-DG}
\end{equation}

\subsection{Boost--gauge commutator and the origin of the \texorpdfstring{\(\Pi\)}{Pi} term}
For one particle,
\(
\mathcal B_i(\mathbf u)=t\,u_\alpha\partial_{i\alpha}+m_i u_\alpha\bar\partial_{i\alpha}
\).
Since \(\mathcal B_i\) has constant coefficients in \(\br_i,\bp_i\), one has
\(G_i(\mathcal B_i^r)=G_i(\mathcal B_i^p)=0\), and therefore \([\,\mathcal B_i,G_i\,]=\mathcal B_i\cdot\nabla_\Gamma\,G_i\).
The \(r\)-component is immediate:
\[
([\,\mathcal B_i,G_i\,])^r_\alpha
=\mathcal B_i\big(\bm\epsilon_\alpha(\br_i)\big)
=t\,u_\beta\partial_{i\beta}\bm\epsilon_\alpha(\br_i)
=\big[t(\mathbf u\!\cdot\!\nabla)\bm\epsilon\big]_\alpha.
\]
For the \(p\)-component, with \(G_i^p{}_\alpha=-\partial_{i\beta}\bm\epsilon_\alpha\,p_{i\beta}\),
\begin{align*}
([\,\mathcal B_i,G_i\,])^p_\alpha
&=\mathcal B_i(G_i^p{}_\alpha) \\
&=t\,u_\gamma\partial_{i\gamma}\!\Big(-\partial_{i\beta}\bm\epsilon_\alpha\,p_{i\beta}\Big)
+m_i u_\gamma\bar\partial_{i\gamma}\!\Big(-\partial_{i\beta}\bm\epsilon_\alpha\,p_{i\beta}\Big) \\
&= -\,\partial_{i\beta}\Big(t\,u_\gamma\partial_{i\gamma}\bm\epsilon_\alpha\Big)\,p_{i\beta}
\;-\;(\partial_{i\gamma}\bm\epsilon_\alpha)\,(m_i u_\gamma).
\end{align*}
The first term is exactly the cotangent-lift \(p\)-component of \(G_i[t(\mathbf u\!\cdot\!\nabla)\bm\epsilon]\).
The second term is independent of \(\bp_i\) and thus is a pure momentum-fiber translation:
\begin{equation}
\begin{aligned}
-(\partial_{i\gamma}\bm\epsilon_\alpha)
(m_i u_\gamma)\,\bar\partial_{i\alpha}
&=\Pi_i[-\bm\chi_{\bm\epsilon,\mathbf u}],
\\
\chi_{\bm\epsilon,\mathbf u,\alpha}(\br_i)
&\equiv m_i u_\gamma
\,\partial_\gamma\bm\epsilon_\alpha(\br_i).
\end{aligned}
\end{equation}

Summing over \(i\) gives
\begin{equation}
[\mathcal B(\mathbf u),G[\bm\epsilon]]
=G\!\big[t(\mathbf u\!\cdot\!\nabla)\bm\epsilon\big]+\Pi\!\big[-\bm\chi_{\bm\epsilon,\mathbf u}\big],
\label{eq:app-comm-BG}
\end{equation}
which exhibits explicitly that the extra \(\Pi\) term originates from the momentum-shift
part \(m_i\mathbf u\!\cdot\!\nabla_{\bp_i}\) of the boost generator.

\subsection{Gauge--gauge commutator}
Let \(\bm\epsilon_1,\bm\epsilon_2\) be smooth spatial vector fields.
For one particle, compute the \(r\)-component:
\begin{align*}
([G_i[\bm\epsilon_1],G_i[\bm\epsilon_2]])^r_\alpha
&=G_i[\bm\epsilon_1]\big(\bm\epsilon_{2,\alpha}(\br_i)\big)-G_i[\bm\epsilon_2]\big(\bm\epsilon_{1,\alpha}(\br_i)\big) \\
&=\bm\epsilon_{1,\beta}\partial_{i\beta}\bm\epsilon_{2,\alpha}
-\bm\epsilon_{2,\beta}\partial_{i\beta}\bm\epsilon_{1,\alpha}
=[\bm\epsilon_1,\bm\epsilon_2]_\alpha,
\end{align*}
which is the spatial Lie bracket \eqref{eq:app-spatialbracket}.
A straightforward (but slightly longer) component calculation for the \(p\)-component yields
\[
([G_i[\bm\epsilon_1],G_i[\bm\epsilon_2]])^p_\alpha
=-\partial_{i\beta}\big([\bm\epsilon_1,\bm\epsilon_2]_\alpha\big)\,p_{i\beta},
\]
so the commutator retains the cotangent-lift structure. Summing over \(i\) gives
\begin{equation}
[G[\bm\epsilon_1],G[\bm\epsilon_2]]
=G\!\big[[\bm\epsilon_1,\bm\epsilon_2]\big].
\label{eq:app-comm-GG}
\end{equation}

\subsection{Momentum-fiber translation--gauge commutator}
Let \(\Pi_i[\bm\chi]=\chi_\alpha(\br_i)\bar\partial_{i\alpha}\).
The \(r\)-component of \([\,\Pi_i[\bm\chi],G_i[\bm\epsilon]\,]\) vanishes because
\(\Pi_i\) differentiates only in \(\bp_i\) while \(\bm\epsilon(\br_i)\) depends only on \(\br_i\).
For the \(p\)-component:
\begin{align*}
([\,\Pi_i[\bm\chi],G_i[\bm\epsilon]\,])^p_\alpha
&=\Pi_i[\bm\chi]\big(-\partial_{i\beta}\bm\epsilon_\alpha\,p_{i\beta}\big)
-G_i[\bm\epsilon]\big(\chi_\alpha(\br_i)\big) \\
&= -(\partial_{i\beta}\bm\epsilon_\alpha)\chi_\beta
-\bm\epsilon_\beta\partial_{i\beta}\chi_\alpha.
\end{align*}
Therefore
\begin{equation}
[\Pi[\bm\chi],G[\bm\epsilon]]
=\Pi\!\Big[-(\bm\epsilon\!\cdot\!\nabla)\bm\chi-(\nabla\bm\epsilon)\bm\chi\Big],
\label{eq:app-comm-PiG}
\end{equation}
which is Eq.~\eqref{eq:comm-PiG} in the main text.

\subsection{Time translation--gauge commutator}
If \(\bm\epsilon=\bm\epsilon(\br)\) is time independent, then clearly
\([t(\tau),G[\bm\epsilon]]=0\).
If \(\bm\epsilon=\bm\epsilon(t,\br)\) depends explicitly on time and
\(t(\tau)=\tau\partial_t\), then \(t(\tau)\) acts only on the coefficients of \(G[\bm\epsilon]\),
yielding
\begin{equation}
[t(\tau),G[\bm\epsilon]]=G[\tau\,\partial_t\bm\epsilon],
\label{eq:app-comm-tG}
\end{equation}
which reproduces Eq.~\eqref{eq:comm-tG}.

\section{Wigner--Eckart--Ward reduction}
\label{app:wew}

This Appendix shows the intermediate steps behind the Wigner--Eckart--Ward reduction used in
Sec.~\ref{subsec:wew}.  We assume a homogeneous isotropic bulk in $n=3$, so the angular dependence is
organized by $SO(3)$ irreducible tensors, spherical harmonics $Y_{\ell m}$, and Clebsch--Gordan (CG)
coefficients.  When CG coefficients are written, the vector index is most naturally interpreted in a
spherical basis $\mu=-1,0,1$; conversion to Cartesian components is a fixed unitary map and does not
affect any radial spectra.
Throughout we use the standard Condon--Shortley normalization
$\int \dd\Omega\,Y_{LM}^*(\hat{\bk})\,Y_{L'M'}(\hat{\bk})=\delta_{LL'}\delta_{MM'}$,
and we write $\int \dd\Omega/(4\pi)$ for the corresponding solid-angle average.

The microscopic density and the universal gauge one-body field are given by
\begin{equation}
\hat\rho(\br)=\sum_{i=1}^N\delta(\br-\br_i),
\qquad
\mathbf J_G(\br)= -\,\kB T\,\nabla\hat\rho(\br).
\end{equation}
With the Fourier convention $\hat\rho(\bk)=\int \dd^3r\,e^{-\mathrm{i}\bk\cdot\br}\hat\rho(\br)$,
\begin{align}
\mathbf J_G(\bk)
&= \int \dd^3r\,e^{-\mathrm{i}\bk\cdot\br}\mathbf J_G(\br)
=\,-\mathrm{i}\,\kB T\,\bk\,\hat\rho(\bk),
\label{eq:app-JGk-IBP}
\end{align}
where the surface term vanishes for periodic boundaries (or for sufficiently fast decay in an unbounded
domain).  Linearity of the ensemble average then gives, for any configuration-only irreducible tensor
insertion $T^{(\ell)}(\mathbf 0)$,
\begin{equation}
\big\langle T^{(\ell)}(\mathbf 0)\otimes \mathbf J_G(\bk)\big\rangle
=\,-\mathrm{i}\,\kB T\,k\,\hat{\bk}\,
\big\langle T^{(\ell)}(\mathbf 0)\,\hat\rho(\bk)\big\rangle .
\label{eq:app-wew-long-rho}
\end{equation}
In a translationally invariant bulk,
$\langle \hat\rho(\bk)\rangle=(2\pi)^3\rho\,\delta(\bk)$, where $\delta(\bk)$ is the
three-dimensional Dirac delta distribution.
Thus for $\bk\neq\mathbf 0$ one has $\langle\hat\rho(\bk)\rangle=0$ and therefore
$\hat\rho(\bk)=\delta\hat\rho(\bk)$ inside correlators.
  At $\bk=\mathbf 0$ the left-hand side vanishes
because $\mathbf J_G(\mathbf 0)=\mathbf 0$ from  Eq.~\eqref{eq:app-JGk-IBP}.  Hence
Eq.~\eqref{eq:app-wew-long-rho} is equivalent to Eq.~\eqref{eq:wew-long},
\begin{equation}
\big\langle T^{(\ell)}(\mathbf 0)\otimes \mathbf J_G(\bk)\big\rangle
=\,-\mathrm{i}\,\kB T\,k\,\hat{\bk}\,
\big\langle T^{(\ell)}(\mathbf 0)\,\delta\hat\rho(\bk)\big\rangle .
\end{equation}
The longitudinal character is explicit in components (spherical basis $\mu=-1,0,1$):
\[
\big\langle T^{(\ell)}_{m}(\mathbf 0)\,J_{G,\mu}(\bk)\big\rangle
=-\,\mathrm{i}\,\kB T\,k\,\hat k_\mu\,
\big\langle T^{(\ell)}_{m}(\mathbf 0)\,\delta\hat\rho(\bk)\big\rangle ,
\]
so the correlator has no transverse projection.

Isotropy implies that the rank-$\ell$ object
$\langle T^{(\ell)}_{m}(\mathbf 0)\,\delta\hat\rho(\bk)\rangle$
is built solely from $\hat{\bk}$ and therefore admits a harmonic expansion
\begin{equation}
\begin{aligned}
\big\langle T^{(\ell)}_{m}(\mathbf 0)\,\delta\hat\rho(\bk)\big\rangle
&=4\pi\sum_{m'=-\ell}^{\ell}\mathcal A^{(\ell)}_{m m'}(k)\,Y_{\ell m'}(\hat{\bk}),
\\
\mathcal A^{(\ell)}_{m m'}(k)
&=\int \frac{\dd\Omega_{\hat{\bk}}}{4\pi}\,
Y_{\ell m'}^*(\hat{\bk})\,
\big\langle T^{(\ell)}_{m}(\mathbf 0)\,\delta\hat\rho(\bk)\big\rangle .
\end{aligned}
\label{eq:app-A}
\end{equation}
In an isotropic bulk, rotational invariance further implies
$\mathcal A^{(\ell)}_{m m'}(k)=\delta_{m m'}\,\mathcal A_\ell(k)$, so each component is proportional to
$Y_{\ell m}(\hat{\bk})$ with a single reduced radial amplitude $\mathcal A_\ell(k)$.

Substitution into Eq.~\eqref{eq:wew-long} reduces the problem to decomposing
$\hat{\bk}\,Y_{\ell m}(\hat{\bk})$ into irreducible $SO(3)$ sectors.  In a spherical basis,
\begin{equation}
\hat k_\mu=\sqrt{\frac{4\pi}{3}}\,Y_{1\mu}(\hat{\bk}),
\qquad \mu=-1,0,1,
\label{eq:app-hatk-sph}
\end{equation}
so products of the form $\hat k_\mu Y_{\ell m}$ reduce to products of two spherical harmonics.
Using the standard product identity
\begin{equation}
\begin{aligned}
Y_{\ell m}(\hat{\bk})\,Y_{1\mu}(\hat{\bk})
&=\sum_{L=\ell-1}^{\ell+1}\sum_{M=-L}^{L}
\sqrt{\frac{(2\ell+1)\,3}{4\pi(2L+1)}} \\
&\quad\times
\langle \ell 0;1 0|L 0\rangle\,
\langle \ell m;1\mu|L M\rangle\,
Y_{L M}(\hat{\bk}) .
\end{aligned}
\label{eq:app-Yprod}
\end{equation}

one obtains explicitly the tensor-product reduction $1\otimes \ell=(\ell-1)\oplus\ell\oplus(\ell+1)$.
In particular, the $L=\ell$ coefficient vanishes identically because
$\langle \ell 0;1 0|\ell 0\rangle=0$ (equivalently $(\ell\,1\,\ell;0\,0\,0)=0$), hence only the
$L=\ell\pm1$ sectors survive in $\hat k_\mu Y_{\ell m}$. This matches the fact that a gradient field
has no toroidal component.
The remaining $L=\ell\pm1$ pieces can therefore be regrouped into two irreducible channels, yielding
the two-channel form shown in Eq.~\eqref{eq:WEW}.

Let $Q_{2m}\equiv T^{(2)}_m$ and define
$C_{2m,i}(\bk)\equiv\langle Q_{2m}(\mathbf 0)\,J_{G,i}(\bk)\rangle$.
From Eq.~\eqref{eq:wew-long} and Eq.~\eqref{eq:app-A} (with $\ell=2$) one may write
\begin{equation}
C_{2m,i}(\bk)
=\,-\mathrm{i}\,\kB T\,k\,\hat k_i\,
4\pi\sum_{m'=-2}^{2}\mathcal A^{(2)}_{m m'}(k)\,Y_{2m'}(\hat{\bk}).
\label{eq:app-Cstart}
\end{equation}
The pair of indices $(m,i)$ lives in the tensor-product space $2\otimes 1=1\oplus 2\oplus 3$.
Longitudinality removes the $L=2$ (toroidal) sector, leaving only $L=1$ and $L=3$.
A convenient basis for these two allowed angular sectors is
\begin{align}
\mathcal V^{(1)}_{m,i}(\hat{\bk})
&\equiv \sum_{M=-1}^{1}\langle 2\,m;1\,i|1\,M\rangle\,Y_{1M}(\hat{\bk}),
\label{eq:app-V1}\\
\mathcal V^{(3)}_{m,i}(\hat{\bk})
&\equiv \sum_{M=-3}^{3}\langle 2\,m;1\,i|3\,M\rangle\,Y_{3M}(\hat{\bk}),
\label{eq:app-V3}
\end{align}
which are the unique irreducible couplings $2\otimes 1\to L$ evaluated on the sphere.
In this basis the correlator has the two-spectrum decomposition
\begin{equation}
C_{2m,i}(\bk)
=\,-\mathrm{i}\,k\,\kB T\Big[
\alpha_1(k)\,\mathcal V^{(1)}_{m,i}(\hat{\bk})
+\alpha_3(k)\,\mathcal V^{(3)}_{m,i}(\hat{\bk})
\Big].
\label{eq:app-l2}
\end{equation}
Equation~\eqref{eq:app-l2} is the explicit $SO(3)$ form of Eq.~\eqref{eq:l2}
(in a spherical basis for the vector index).

Introduce the inner product on the sphere (including sums over discrete indices)
\begin{equation}
(\mathcal U,\mathcal W)\equiv
\sum_{m,i}\int\frac{\dd\Omega_{\hat{\bk}}}{4\pi}\,
\mathcal U_{m,i}(\hat{\bk})\,\mathcal W^{*}_{m,i}(\hat{\bk}),
\end{equation}
and define the channel normalizations by
\begin{equation}
\mathcal N_L\equiv(\mathcal V^{(L)},\mathcal V^{(L)})
=\sum_{m,i}\int\frac{\dd\Omega_{\hat{\bk}}}{4\pi}\,
\mathcal V^{(L)}_{m,i}(\hat{\bk})\,\mathcal V^{(L)\,*}_{m,i}(\hat{\bk}),
\qquad L=1,3,
\end{equation}
(so $\mathcal N_1=3/(4\pi)$ and $\mathcal N_3=7/(4\pi)$ for the standard $Y_{LM}$ normalization).
Orthogonality of irreducible sectors implies $(\mathcal V^{(1)},\mathcal V^{(3)})=0$.
Multiplying \eqref{eq:app-l2} by $\mathcal V^{(1)\,*}$, integrating, and using orthogonality yields
\begin{equation}
\sum_{m,i}\int\frac{\dd\Omega_{\hat{\bk}}}{4\pi}\,
\mathcal V^{(1)\,*}_{m,i}(\hat{\bk})\,C_{2m,i}(\bk)
= -\,\mathrm{i}\,k\,\kB T\,\alpha_1(k)\,\mathcal N_1.
\end{equation}

hence
\begin{equation}
\alpha_1(k)
=\frac{\mathrm{i}}{k\,\kB T}\,\frac{1}{\mathcal N_1}
\sum_{m,i}\int\frac{\dd\Omega_{\hat{\bk}}}{4\pi}\,
\mathcal V^{(1)\,*}_{m,i}(\hat{\bk})\,C_{2m,i}(\bk),
\end{equation}
which reproduces Eq.~\eqref{eq:alpha1}.  The same steps with $\mathcal V^{(3)\,*}$ give
\begin{equation}
\alpha_3(k)
=\frac{\mathrm{i}}{k\,\kB T}\,\frac{1}{\mathcal N_3}
\sum_{m,i}\int\frac{\dd\Omega_{\hat{\bk}}}{4\pi}\,
\mathcal V^{(3)\,*}_{m,i}(\hat{\bk})\,C_{2m,i}(\bk),
\end{equation}
which is Eq.~\eqref{eq:alpha3}.  Finally, $\mathbf J_G(\bk)\propto \bk$ implies
$C_{2m,i}(\bk)=\mathcal O(k)$ as $k\to0$, so both spectra $\alpha_1(k)$ and $\alpha_3(k)$ remain finite in
the long-wavelength limit.

\end{document}